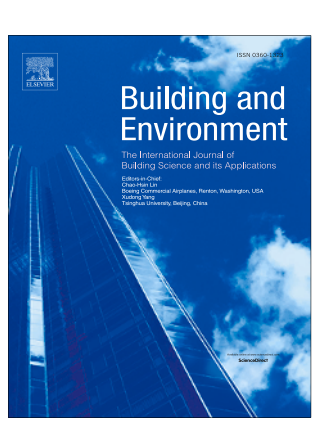

# Heating in human-HVAC interaction for smart homes: An interdisciplinary overview

Delong Korus-Du[a,*], Gunnar Stevens[a], Alexander Boden[b], Lenneke Kuijer[c], Apostolos K. Vavouris[d], Md Shajalal[b], Philip Engelbutzeder[a], Omid Veisi[e], Peter Tolmie[a]

[a] Department of Human-Computer Interaction, University of Siegen, Siegen, Germany
[b] Fraunhofer Institute for Applied Information Technology FIT, Sankt Augustin, Germany
[c] Department of Industrial Design, Eindhoven University of Technology, Eindhoven, Netherlands
[d] Department of Electrical Engineering, University of Strathclyde, Glasgow, United Kingdom
[e] Faculty of Information, University of Toronto, Toronto, Canada

## HIGHLIGHTS

- Bridges fragmented smart-home heating research from 541 studies across 7 discipline.
- Links Social Residents' Experience and Practices with Heating in HVAC System Mechanics.
- Defines H-HVAC Interaction control and feedback as Situated Interaction Dynamics.
- Identifies four tensions between technical performance and household outcomes.
- Guides smart-home heating design for health, affordability, and sustainability.

## ARTICLE INFO

## ABSTRACT

As part of HVAC systems, residential heating provides foundational infrastructure for human habitation in cold weather. However, research on how residents interact with HVAC systems, particularly heating systems, remains fragmented across architecture, engineering, informatics, physiology, psychology, sociology, and design. Based on 541 studies from these fields, this review integrates interdisciplinary research on Heating in Human–HVAC Interaction in smart homes.

The resulting synthesis is conceptualized through the Situated Interaction Dynamics of control and feedback between users and systems. User-initiated interactions involve monitoring past and present system performance and planning future operation, while system-initiated interactions rely on sensor networks to trigger automation or provide information enabling user action.

These interaction dynamics connect Residents' Experience and Practices with Heating in HVAC System Mechanics. Residents' Experience and Practices include thermal comfort and energy management, where thermal comfort involves both individual physiological and psychological experiences of indoor climate and social practices shaped by norms, empathy, and negotiation among cohabitants. Heating in HVAC System Mechanics includes thermal conditions and energy performance. Thermal conditions concern the regulation of air temperature, mean radiant temperature, air velocity, and relative humidity, while energy performance concerns efficiency and environmental impact.

This overview highlights four interdisciplinary tensions: sensed versus lived conditions, personalization versus negotiation, efficiency versus health, and automation versus agency. The resulting framework offers a conceptual lens to interpret heating interactions and design Human–HVAC Interaction that balances IEQ-driven healthy thermal conditions, affordability, and sustainability.

* Corresponding author.
*Email address:* delong.korusdu@uni-siegen.de (D. Korus-Du).

Open access funding enabled and organized by Projekt DEAL.

**Nomenclature**

| | |
|---|---|
| ASHRAE | American Society of Heating, Refrigerating and Air-Conditioning Engineers |
| COP | Coefficient of Performance |
| CPCI-S | Conference Proceedings Citation Index–Science |
| DOI | Digital Object Identifier |
| HBI | Human–Building Interaction |
| HCI | Human–Computer Interaction |
| HEMS | Home Energy Management System |
| HVAC | Heating, Ventilation, and Air Conditioning |
| IAQ | Indoor Air Quality |
| IEEE | Institute of Electrical and Electronics Engineers |
| IEQ | Indoor Environmental Quality |
| IoT | Internet of Things |
| PMV | Predicted Mean Vote |
| PPD | Predicted Percentage of Dissatisfied |
| RH | Relative Humidity |
| RQ | Research Question |
| SEMS | Smart Energy Management System |
| SSCI | Social Sciences Citation Index |
| TCPS | Thermal Comfort Prediction Systems |
| TI | Title field in Web of Science search syntax |
| VOCs | Volatile Organic Compounds |

## 1. Introduction

### 1.1. Background: residential heating in human-building interaction

As societal change has resulted in people spending more time indoors since the mid-20th century [1], residential heating has come to weigh significantly on energy demand, environmental impact, household expenditure, and everyday experiences of indoor conditions [2–6]. Government energy reports indicate that buildings in polar, subpolar, and temperate zones account for approximately 30% of final energy use, with nearly half devoted to heating and approximately 2.4 Gt of direct and 1.7 Gt of indirect $CO_2$ emissions [2–5]. Although more efficient and lower-carbon heating technologies are increasingly available, fossil fuels continue to supply over 60% of heating demand [2–4]. As geopolitical tensions, expanding digital infrastructures, and increasing electricity demand place additional pressure on energy systems [7–9], families and communities face growing concerns about the affordability of domestic energy [10].

Residential heating is not only a technical energy system, but also shapes what residents experience, interpret, and manage in everyday life [11–14]. Before central heating, residents relied more extensively on clothing, localised heat sources, shared warm rooms, spatial routines, and seasonal adaptation, whereas the expansion of central heating contributed to expectations of more stable and spatially uniform indoor temperatures [11,12,14]. Architectural developments likewise reflected changing relationships among health, heating, ventilation, and building design. From the late 20th century, climate research and environmental policy increased attention to the energy and environmental performance of buildings [15]. Clean-air legislation and international agreements such as the Montreal Protocol also made the environmental effects of energy and refrigerant technologies increasingly visible within building and HVAC policy [16,17]. Passive heating and natural ventilation, biomass systems, district heating, and other lower-impact building systems illustrate continuing efforts to combine appropriate indoor conditions with greater energy efficiency and lower environmental impact [18–21].

Digitisation subsequently transformed heating into a programmable, sensor-based, and automated infrastructure. Ubiquitous and context-aware computing established a technical basis for embedding sensing and computation within everyday environments [22,23]. Smart homes subsequently became an important site for connecting domestic technologies with comfort, health, convenience, security, communication, and energy management [24,25]. Connected heating systems introduced sensing, scheduling, feedback, remote control, and automated actuation, bringing heating closer to research on human–building interaction [26–28], human–machine interaction [29], and HCI in smart homes [30,31]. These developments position residential heating as an important area within worldwide efforts associated with the United Nations' Sustainable Development Goals [32,33]. Heating is connected not only to energy efficiency and decarbonisation but also to health protection, domestic affordability, and residents' ability to understand and influence the environments in which they live [10,34,35]. Thermal discomfort and inadequate indoor temperatures may pose particular health risks to vulnerable populations [34], especially during periods of substantial seasonal temperature variation.

### 1.2. State of the art: fragmented perspectives on smart residential heating

Existing research has achieved significant progress in the development and study of residential heating systems, however, smart homes do not present a technological silver bullet for sustainable or comfortable heating. Over-reliance on cybernetic automation may fail to sustain long-term behavioural change, increase the effort that residents devote to household care, and disrupt established habits of domestic technology use [13,36,37]. These difficulties may contribute to information avoidance and the abandonment of monitoring or maintenance over time [36,37]. Smart devices are also constrained by the partiality of environmental sensing, occupancy inference, comfort modelling, and computational representations of residents' needs [38–40]. When measurements or inferred preferences do not correspond to residents' bodily experiences or household situations, automated control may produce inappropriate or difficult-to-correct outcomes [39–41].

In parallel, consumer and social research grounded in Social Practice Theory describes how materials, competences, and meanings constitute domestic heating practices. This research connects heating to everyday routines, community learning, social norms, housing infrastructures, and changing expectations of indoor comfort [11–14]. Empirical research demonstrates that household routines and practices substantially shape energy consumption, highlighting the limitations of focusing solely on technical system efficiency [13,42]. Users with varying levels of energy literacy may also face difficulties in adopting, understanding, configuring, and maintaining smart heating systems [36,37,43]. Moreover, infrastructural improvements do not translate directly into proportional energy reductions because household energy demand may be affected by rebound and prebound effects [44–47].

Sustainable Interaction Design and Sustainable HCI have similarly proposed interfaces and interactive systems that aim to foster energy literacy, reflection, and conservation throughout the lifecycle of domestic technologies [48–50]. This research has long argued that technologies for everyday life must be understood as elements of socio-technical practices rather than as purely technical optimisation problems [14,41,49,50]. Energy management nevertheless requires feedback that residents can understand, control that they can meaningfully exercise, and automation that remains visible and correctable [35,51].

Human–Building Interaction (HBI) provides another important foundation by examining relationships among people, artefacts, computational systems, and built environments [28,52,53]. HBI foregrounds interaction with built environments, whereas HCI has historically focused more directly on computational artefacts [54]. Early HCI often framed interaction as input and output between users and computers [55–57], whereas later research emphasized situated and embodied interaction, as well as the mental models through which people understand technologies and their effects [58–60]. This shift is particularly important for heating because computational actions alter material, environmental, and bodily conditions rather than only an on-screen state.

Human interaction with heating is therefore not simply an exchange between one user and one thermostat. Residents use settings, clothing, windows, rooms, schedules, and overrides, while systems sense environmental conditions, construct models, infer demand, and actuate heating equipment [13,38,51]. This resembles a *Sense–Plan/Think–Act* cycle [60–63], but residents also draw on bodily sensation, memory, expectation, health, cost, and social context [64–66]. Sensory ethnography and social-practice research similarly connect bodily sensation, behaviour, domestic routines, and consumption [12,64–66].

Smart homes further introduce privacy, cybersecurity, maintenance, connectivity, and outage risks [67,68]. Continuous data collection and failures in connectivity or power may disrupt domestic routines and complicate responsibility for correction and repair [67,68]. Over-automation may weaken residents' agency and fail to accommodate the relational and socio-cultural meanings of home [50,69,70]. Although smart heating may reduce operating energy, its digital infrastructures also produce environmental impacts through computation, manufacturing, replacement, and disposal [48,71]. Upfront costs, maintenance requirements, renewable-energy variability, and rebound effects further complicate the contribution of smart technologies to sustainable domestic heating [35,44,72–75].

### 1.3. Research gap and AIM of the paper

Despite progress across these areas, research on residential heating remains dispersed across various disciplines. Each field explains an important part of the relationship between people and heating systems, but their analytical objects, conceptual vocabularies, and forms of evidence remain only partially connected. Thermal comfort and occupant-centric control research provide models of environmental conditions, personal variables, system operation, and control performance [76–78]. HBI and HCI explain interfaces, feedback, automation, situated action, and residents' understandings of computational systems [26,28,41,58]. Social-practice research explains how heating becomes embedded in routines, materials, competences, social meanings, and social expectations [11–13]. Sustainable HCI examines energy literacy, intervention, rebound, and the environmental consequences of digital technologies [48–50]. However, no single perspective fully explains how sensing, modelling, actuation, building response, bodily experience, household negotiation, correction, and maintenance become connected within everyday residential heating.

The absence of an integrated overview makes it difficult for designers, engineers, and researchers to identify how residents' experiences and practices [12–14,31,42,48–50,79], thermal comfort research, and HVAC system mechanics [5,39,40,76,77,80,81] jointly shape interactions with smart heating technologies. This fragmentation also obscures tensions between automated control and resident agency, standardized models and lived comfort, individualized preferences and shared household practices, energy savings and health protection, and short-term efficiency and wider environmental impact.

This paper therefore introduces **Human–HVAC Interaction** as a multidisciplinary analytical lens for examining smart residential heating. The term does not replace existing concepts such as Human–Building Interaction [26,82], occupant-centric control [83], adaptive thermal comfort [78,84], or Social Practice Theory [12,14]. Instead, it connects their otherwise fragmented contributions by examining the interaction through which residents and heating systems mutually shape indoor thermal conditions and domestic energy practices.

The scope of the review is specifically *heating-focused Human–HVAC Interaction in smart residential environments*. The review does not attempt to provide a complete account of cooling, non-residential HVAC, or all four dimensions of IEQ. Rather, it examines heating as a socio-technical interaction involving computational control, HVAC mechanics, building response, bodily and subjective experience, everyday practice, household relationships, affordability, and environmental impact.

By adopting a scoping review methodology, the paper addresses the following research questions:

1. **Understanding Human–HVAC Interaction:** How is heating-focused Human–HVAC Interaction conceptualized and studied across relevant disciplines in smart residential environments?
2. **Designing H–HVAC Interaction:** What interdisciplinary tensions, research gaps, and design opportunities emerge when residents' experiences and practices are examined together with HVAC system mechanics?

Building on 541 studies, this review contributes an integrated overview that makes interdependencies across disciplinary domains visible. As shown in Fig. 1, heating-focused Human–HVAC Interaction comprises situated interaction dynamics connecting residents' experiences and practices with HVAC system mechanics. These interactions unfold through sensing, modelling, actuation, building response, bodily experience, household negotiation, correction, and maintenance.

The contribution of the framework is threefold. First, it defines the analytical scope of heating-focused Human–HVAC Interaction by connecting technical, bodily, behavioural, and social processes that are commonly studied separately. Second, it organises the literature through three interdependent areas: situated interaction dynamics, residents' experiences and practices, and HVAC system mechanics. Third, it identifies interdisciplinary tensions that must be addressed when designing residential heating systems intended to support healthy thermal conditions, affordability, and sustainability. The framework therefore provides a conceptual lens through which researchers can locate their contributions, examine the assumptions and limitations of discipline-specific representations, and connect them to the wider interaction cycle of residential heating.

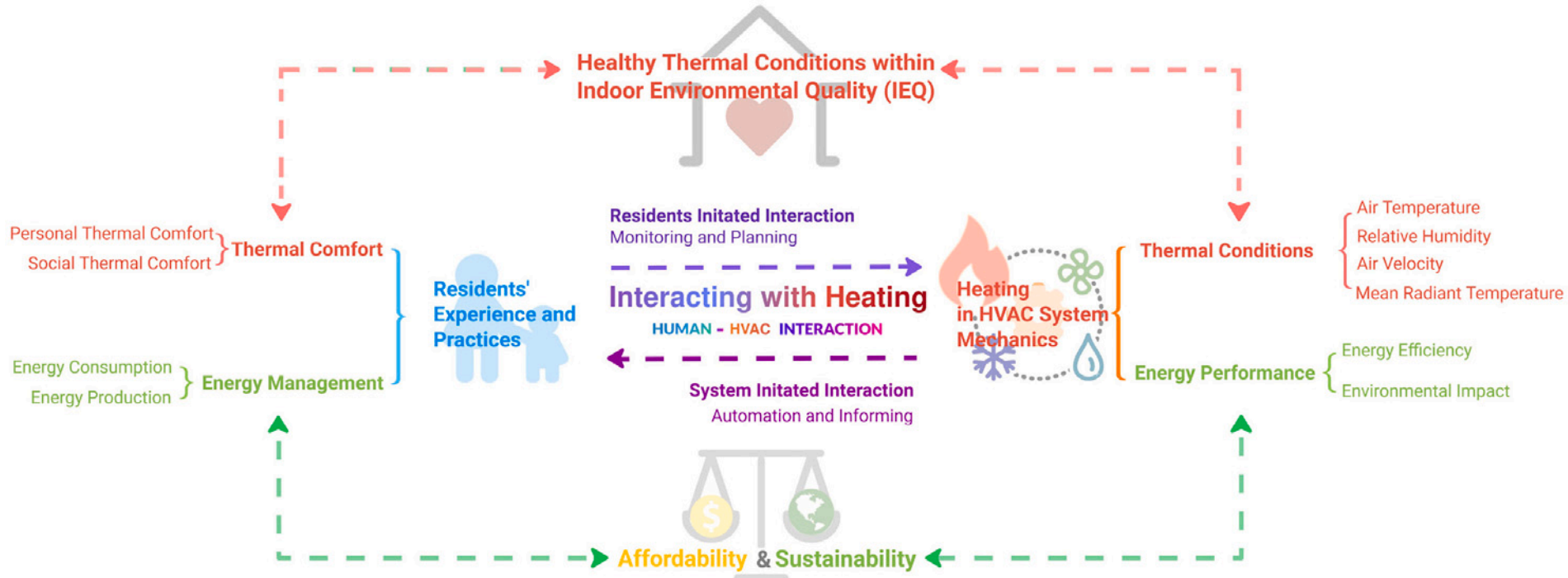

**Fig. 1.** Overview of heating-focused human–HVAC interaction in smart residential environments.

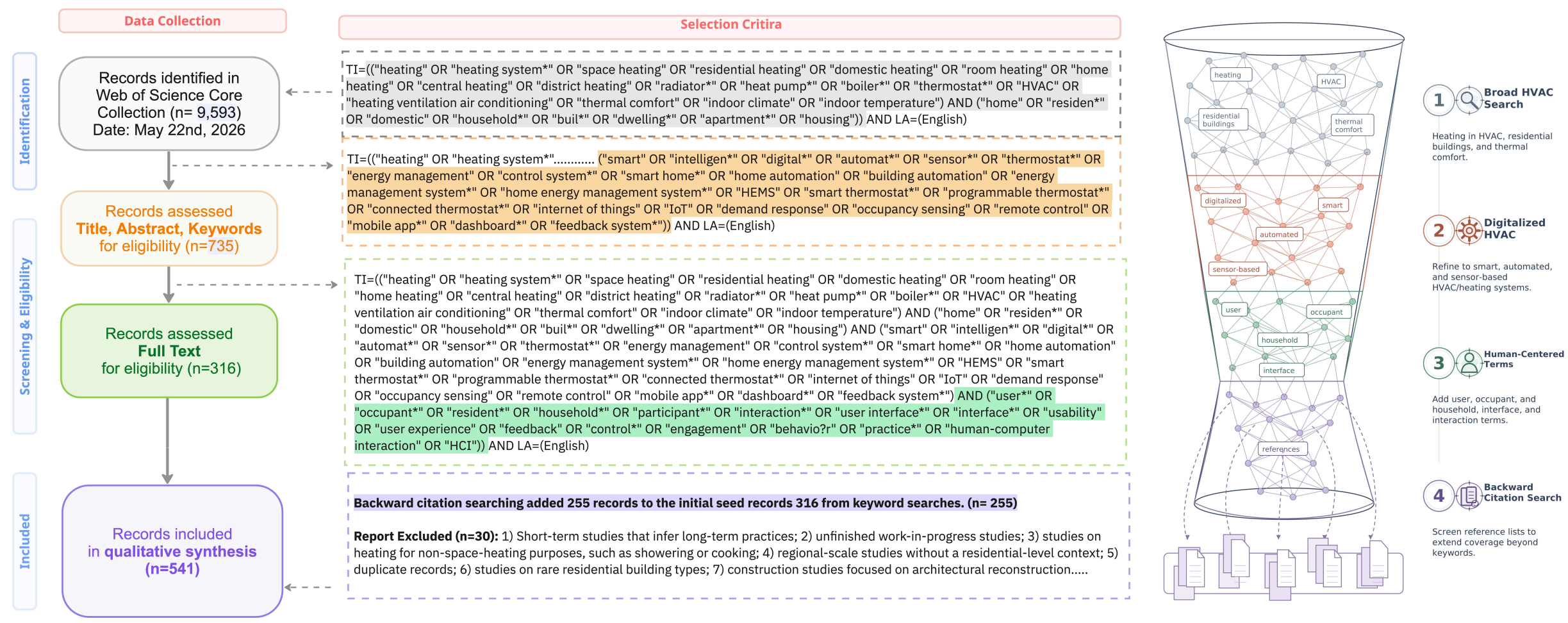

**Fig. 2.** PRISMA-informed flow of identification, screening, backward citation searching, and inclusion.

## 2. Method

### 2.1. Data collection

Our literature search was informed by PRISMA review protocols [85] (Fig. 2). The review focused on Human–HVAC Interaction in smart residential heating, including ventilation and indoor-air-quality processes that interacted directly with heating. It was not designed as a comprehensive review of cooling, non-residential HVAC, or all four IEQ domains.

The search was completed on May 22, 2026, with no lower publication-date restriction and was limited to English-language records. Web of Science Core Collection was the primary structured source, covering SCI-EXPANDED, SSCI, A&HCI, and CPCI-S. Scopus, IEEE Xplore, and ACM Digital Library were used as supplementary coverage checks. Comparison showed substantial overlap with Web of Science, while Scopus returned fewer additional records and IEEE Xplore and ACM Digital Library contributed smaller specialised sets. They were therefore used to identify otherwise missing eligible studies, not as separate PRISMA identification streams. Database-specific adaptations and BibTeX exports are supplied in the supplementary materials.

As an initial scoping step, we examined the breadth of heating-, HVAC-, building-, and smart-system literature. This broad search yielded 13,261 records and demonstrated that unrestricted retrieval was too heterogeneous for the heating-specific scope. A narrower search produced a 9593-record starting set. The formal retrieval then used staged title searches (TI), followed by title–abstract–keyword assessment. Title searching increased precision but could miss relevant studies using disciplinary rather than heating or interaction terminology; supplementary database checks and backward citation searching were used to mitigate, but not eliminate, this limitation.

The search proceeded in four stages, progressively narrowing the literature from residential heating and HVAC research towards studies relevant to Heating in Human–HVAC Interaction.

**Search Stage 1: Broad HVAC search for heating within residential and building contexts.** The first stage identified heating or HVAC studies in residential, domestic, household, dwelling, apartment, housing, or building contexts. It included space, central, and district heating, radiators, heat pumps, boilers, thermal comfort, and indoor-climate terms, producing a 9593 record starting set.

**Search Stage 2: Digitalised, smart, automated, and sensor-based HVAC/heating systems.** The second stage added terms for automation, sensing, thermostats, energy-management and control systems, smart homes, IoT, demand response, occupancy sensing, applications, dashboards, and feedback. This refinement produced 735 records for title–abstract–keyword assessment.

**Search Stage 3: Human-interaction-oriented heating/HVAC research.** The third stage added terms for users, occupants, residents, households, participants, interaction, interfaces, usability, user experience, feedback, control, engagement, behaviour, practices, and HCI. After title–abstract–keyword assessment, 316 records proceeded to full-text review. The complete Web of Science query and equivalent adaptations for Scopus, IEEE Xplore, and ACM Digital Library are supplied in the supplementary search file. Records were deduplicated using DOIs where available, followed by manual comparison of title, author, and publication year.

**Search Stage 4: Full-text screening and backward citation searching.** Studies were eligible when they examined heating in residential HVAC, smart-home, or domestic-energy contexts; involved residents, occupants, households, users, or participants; and addressed sensing, control, feedback, automation, interfaces, comfort practices, energy management, maintenance, or decision-making. Studies were excluded when they addressed only material performance, combustion chemistry, industrial heating, construction, regional modelling without a residential context, non-space-heating applications, or non-residential systems without a transferable interaction contribution. Full-text exclusions also covered insufficient methods or findings, unsupported inferences about long-term household practice, duplicates discovered during reconciliation, and highly atypical residential cases with limited transferability.

Of the 316 full texts assessed, 30 were excluded for one or more of these reasons, leaving 286 database-derived studies. Reason-specific counts were not reconstructed when a report met multiple exclusion categories.

Backward citation searching was used because relevant research is distributed across building science, thermal comfort, engineering, informatics, HCI, household-energy research, environmental psychology, sociology, and social-practice research, where similar problems are described through different terminology [86]. Connected Papers served as a navigation aid for selecting seeds across the main disciplinary and thematic clusters. Seeds combined early foundational, highly cited, and recent publications; citation counts identified influential work but were not an eligibility criterion. Reference lists were screened using the same eligibility criteria, and newly included studies could seed another round. Searching stopped when one complete additional round produced no eligible studies. All candidates were checked against the existing library before inclusion.

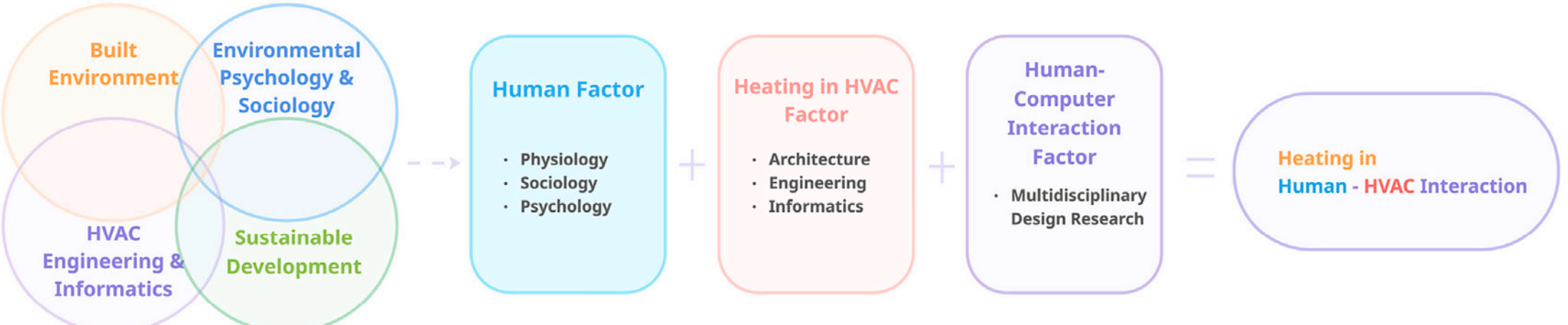

**Fig. 3.** Categorising heating in human–HVAC interaction studies into related fields.

This process identified 255 additional eligible studies. Combined with the 286 database-derived studies, the final corpus comprised 541 unique publications. Potential citation-network bias was mitigated by selecting seeds across disciplines, periods, and citation profiles, but it remains a review limitation.

### 2.2. Data analysis

Our data analysis and overview development followed the original three-stage process.

**Step 1: Mapping the field.** As shown in Fig. 3, the 541 publications originated from four broad research areas: the built environment, environmental psychology and sociology, HVAC engineering and informatics, and sustainable development. Each publication could be mapped to more than one discipline because many crossed physiological, psychological, social, engineering, computing, architectural, health, and energy perspectives. The first author served as the primary analyst and maintained the study-level evidence matrix and codebook. Other authors re-read literature relevant to their expertise and reviewed disciplinary assignments.

**Step 2: Conceptual abstraction.** We conducted thematic analysis across the publications [87,88]. For each study, the evidence matrix recorded bibliographic identity and provenance, disciplinary affiliation, objective, context and population, heating or HVAC technology, method, primary findings and outcomes, and concepts relevant to Human–HVAC Interaction. Titles, abstracts, and keywords supported initial inductive coding, followed by full-text reading to refine each study's contribution and evidential scope.

Recurring codes included thermal experience and perception, household practices and negotiation, sensing and prediction, control and automation, feedback and interfaces, building and HVAC mechanics, affordability and energy poverty, energy management, maintenance, health and vulnerability, and sustainability. The codebook was iteratively revised when studies exposed missing or overlapping definitions. Multi-label coding was retained rather than forcing each publication into one discipline or theme. When interpretations differed, the relevant full text was re-read and discussed until an agreed interpretation or explicit multi-category assignment was retained.

**Step 3: Conceptual organisation and interpretation.** Codes were compared across publications, clustered into higher-order categories, and examined for recurring relationships, contradictions, and gaps among technical evidence, resident experience, household organisation, and energy outcomes. Following Jabareen's understanding of conceptual frameworks as "networks of concepts" [89], the categories were organised into three interdependent dimensions: *Residents' Experience and Practices*, *Heating and HVAC System Mechanics*, and *Situated Interaction Dynamics*. The synthesis also generated the coupled objectives and interdisciplinary tensions examined in the Discussion.

Following the concept-centric approach of Webster and Watson [90], the corpus was interpreted through constructs rather than sequential author summaries. For example, an override could indicate control error, discomfort, repair of opaque automation, household negotiation, or cost containment, depending on the evidence available. The analysis therefore asked what evidence distinguished or connected these explanations rather than selecting one universal interpretation. The Results section reports themes and representative references. The supplementary matrix and codebook materials are available in the project repository: https://github.com/delongkd/Heating-In-Human-HVAC-Interaction-An-Interdisplinary-Review.

Across these steps, the analysis developed over three and a half years through sustained literature engagement and discussions with coauthors, scholars, domain experts, and HVAC engineers in Europe and the United States, including through the Marie Curie network and the ASHRAE community. These exchanges informed the interpretation and prompted revision where needed.

## 3. Results

The results are synthesised into three themes: User Experience, Heating in HVAC Mechanics, and Situated Interaction Dynamics.

### 3.1. Residents' experience & practices – personal and social thermal comfort of IEQ

*Personal thermal comfort.* Personal thermal comfort refers to an individual's subjective evaluation and acceptance of a thermal environment [81,91]. It forms the thermal component of the four-domain concept of indoor environmental quality (IEQ), which also encompasses indoor air quality, acoustic conditions, and visual conditions [92,93]. *Indoor environmental quality* (IEQ) refers broadly to thermal, air-quality, acoustic, and visual conditions [92,93]. Thermal comfort is considered with reference to ISO 7730:2025 and ANSI/ASHRAE 55-2023, while ISO 17772-1 and EN 16798-1 provide the broader IEQ framework [80,81,92,93]. Heat-balance assessment conventionally considers four environmental factors: air temperature, mean radiant temperature, air speed, and relative humidity [76,80]. It also considers two personal factors, metabolic rate and clothing insulation [76,81]. These six factors provide inputs to thermal-comfort assessment; additional influences shape how a particular person experiences a room [77,94]. Exposure duration, thermal gradients, radiant asymmetry, health, acclimatization, expectation, and available adaptive actions can alter the relationship between measured conditions and reported comfort [95]. Air quality is therefore assessed separately from the six thermal-comfort factors through pollutant concentrations, sources, ventilation, and related indicators [92,96]. Likewise, ISO 7730 and ASHRAE 55 are used here to assess thermal conditions within the broader IEQ framework.

*Research on physiological factors* shows that personal thermal comfort is closely linked to thermoregulation. The body regulates heat through vasodilation, vasoconstriction, sweating, shivering, and metabolic heat production [97,98]. The capacity to deploy these responses varies

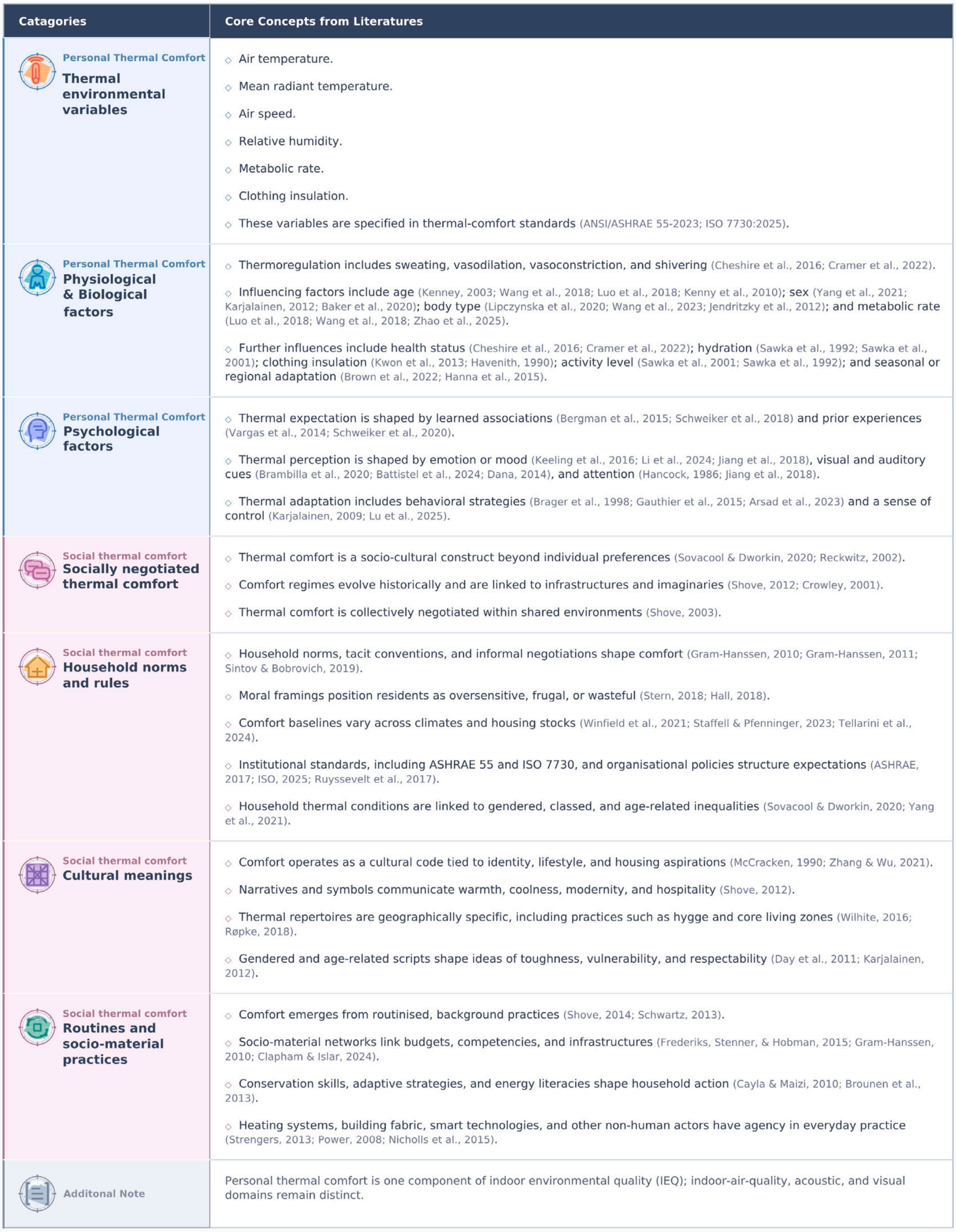
HEATING in HUMAN-HVAC INTERACTION / TABLE 01

## Residents' Personal and Socially Negotiated Thermal Comfort

Personal Thermal Comfort | Social thermal comfort

| Catagories | Core Concepts from Literatures |
|---|---|
| Personal Thermal Comfort<br>**Thermal environmental variables** | ◇ Air temperature.<br>◇ Mean radiant temperature.<br>◇ Air speed.<br>◇ Relative humidity.<br>◇ Metabolic rate.<br>◇ Clothing insulation.<br>◇ These variables are specified in thermal-comfort standards (ANSI/ASHRAE 55-2023; ISO 7730:2025). |
| Personal Thermal Comfort<br>**Physiological & Biological factors** | ◇ Thermoregulation includes sweating, vasodilation, vasoconstriction, and shivering (Cheshire et al., 2016; Cramer et al., 2022).<br>◇ Influencing factors include age (Kenney, 2003; Wang et al., 2018; Luo et al., 2018; Kenny et al., 2010); sex (Yang et al., 2021; Karjalainen, 2012; Baker et al., 2020); body type (Lipczynska et al., 2020; Wang et al., 2023; Jendritzky et al., 2012); and metabolic rate (Luo et al., 2018; Wang et al., 2018; Zhao et al., 2025).<br>◇ Further influences include health status (Cheshire et al., 2016; Cramer et al., 2022); hydration (Sawka et al., 1992; Sawka et al., 2001); clothing insulation (Kwon et al., 2013; Havenith, 1990); activity level (Sawka et al., 2001; Sawka et al., 1992); and seasonal or regional adaptation (Brown et al., 2022; Hanna et al., 2015). |
| Personal Thermal Comfort<br>**Psychological factors** | ◇ Thermal expectation is shaped by learned associations (Bergman et al., 2015; Schweiker et al., 2018) and prior experiences (Vargas et al., 2014; Schweiker et al., 2020).<br>◇ Thermal perception is shaped by emotion or mood (Keeling et al., 2016; Li et al., 2024; Jiang et al., 2018), visual and auditory cues (Brambilla et al., 2020; Battistel et al., 2024; Dana, 2014), and attention (Hancock, 1986; Jiang et al., 2018).<br>◇ Thermal adaptation includes behavioral strategies (Brager et al., 1998; Gauthier et al., 2015; Arsad et al., 2023) and a sense of control (Karjalainen, 2009; Lu et al., 2025). |
| Social thermal comfort<br>**Socially negotiated thermal comfort** | ◇ Thermal comfort is a socio-cultural construct beyond individual preferences (Sovacool & Dworkin, 2020; Reckwitz, 2002).<br>◇ Comfort regimes evolve historically and are linked to infrastructures and imaginaries (Shove, 2012; Crowley, 2001).<br>◇ Thermal comfort is collectively negotiated within shared environments (Shove, 2003). |
| Social thermal comfort<br>**Household norms and rules** | ◇ Household norms, tacit conventions, and informal negotiations shape comfort (Gram-Hanssen, 2010; Gram-Hanssen, 2011; Sintov & Bobrovich, 2019).<br>◇ Moral framings position residents as oversensitive, frugal, or wasteful (Stern, 2018; Hall, 2018).<br>◇ Comfort baselines vary across climates and housing stocks (Winfield et al., 2021; Staffell & Pfenninger, 2023; Tellarini et al., 2024).<br>◇ Institutional standards, including ASHRAE 55 and ISO 7730, and organisational policies structure expectations (ASHRAE, 2017; ISO, 2025; Ruyssevelt et al., 2017).<br>◇ Household thermal conditions are linked to gendered, classed, and age-related inequalities (Sovacool & Dworkin, 2020; Yang et al., 2021). |
| Social thermal comfort<br>**Cultural meanings** | ◇ Comfort operates as a cultural code tied to identity, lifestyle, and housing aspirations (McCracken, 1990; Zhang & Wu, 2021).<br>◇ Narratives and symbols communicate warmth, coolness, modernity, and hospitality (Shove, 2012).<br>◇ Thermal repertoires are geographically specific, including practices such as hygge and core living zones (Wilhite, 2016; Røpke, 2018).<br>◇ Gendered and age-related scripts shape ideas of toughness, vulnerability, and respectability (Day et al., 2011; Karjalainen, 2012). |
| Social thermal comfort<br>**Routines and socio-material practices** | ◇ Comfort emerges from routinised, background practices (Shove, 2014; Schwartz, 2013).<br>◇ Socio-material networks link budgets, competencies, and infrastructures (Frederiks, Stenner, & Hobman, 2015; Gram-Hanssen, 2010; Clapham & Islar, 2024).<br>◇ Conservation skills, adaptive strategies, and energy literacies shape household action (Cayla & Maizi, 2010; Brounen et al., 2013).<br>◇ Heating systems, building fabric, smart technologies, and other non-human actors have agency in everyday practice (Strengers, 2013; Power, 2008; Nicholls et al., 2015). |
| Additonal Note | Personal thermal comfort is one component of indoor environmental quality (IEQ); indoor-air-quality, acoustic, and visual domains remain distinct. |

with age, sex, body composition, activity, health, hydration, clothing, and acclimatization [99,100]. Older adults can experience reduced metabolic heat production, altered vasomotor responses, and impaired sweating [101–103]; chronic cardiovascular, metabolic, or neurological conditions can further restrict thermoregulation [97,104]. Sex-related physiological differences, including differences in metabolic rate and hormonal state, can also influence thermal preference [105–107]. These variations mean that an average comfort prediction does not establish equal thermal protection across a population.

Beyond physiology, research on *psychological factors* shows that thermal experience is shaped by expectations, thermal memory, emotion, attention, cognitive appraisal, and perceived control [95,108,109]. Thermal expectations develop through learned associations [110], prior experience [111], and repeated exposure to particular environments [112]. Domestic spaces, for example, may be associated with warmth and cosiness, whereas institutional settings may be associated with cooler or more tightly controlled conditions. When experienced conditions conflict with these expectations, discomfort may be amplified [95].

Thermal perception is also affected by emotion, mood, task engagement, and multisensory cues [113–115]. Lighting colour can modify thermal judgements in some experimental settings [116], while fire imagery [117], wind or fire sounds [118], and virtual contexts [119,120] can bias perceived warmth or coolness without an equivalent change in room temperature. Evidence for the hue–heat hypothesis (HHH), however, is mixed. Laura et al. [121] found no overall improvement in thermal sensitivity when red–warm and blue–cold lighting were congruent; participants showed heterogeneous, sometimes opposing patterns, and the added colour cue may have increased cognitive load. Bellia et al. [122,123] caution that apparent disagreement among HHH studies may also arise from environmental boundary conditions that are insufficiently justified or not comparable. Tests of the hypothesis should therefore define lighting and thermal exposures within an IEQ framework, using illuminance levels and operative temperatures representative of typical indoor environments. Controlled and comparable conditions are needed to distinguish a hue-related effect from differences in illuminance, thermal exposure, experimental protocol, or context. Mixed findings consequently do not by themselves establish either the validity or invalidity of the HHH. Cross-domain effects may be synergistic or antagonistic, but visual and acoustic conditions should remain analytically distinct from thermal variables. These perceptual findings do not, without residential field evidence, demonstrate durable energy savings or provide a safe substitute for heating.

Finally, *thermal adaptation* connects experience to action. Residents adjust clothing, posture, activity, location, windows, thermostats, and heating practices to restore or maintain comfort [124–126]. These actions are shaped by climate and cultural experience [127], as well as by habits, building fabric, household routines, and access to controls. Perceived control can improve acceptance even before an adjustment changes the physical environment [128], whereas an unavailable, slow, or opaque control can intensify dissatisfaction [129]. Reviews and methodological debate further show that the representation of adaptation affects what a thermal-comfort model can legitimately predict [130,131]. Thermal-comfort studies should therefore distinguish physical state, model output, subjective judgement, adaptive opportunity, and realised action rather than treating any one of them as ground truth.

The apparent disagreement between thermal-physiology, modelling, and experience-centred studies often reflects different evidential objects rather than mutually exclusive findings. Standards for measuring physical quantities establish whether an environmental observation is technically credible [132]; subjective-judgement scales establish how an occupant evaluates that exposure [91]; and personal comfort models estimate an individual's response from selected inputs [133]. These objects can agree, but none substitutes for the others. A calibrated sensor can correctly report air temperature while missing radiant asymmetry or the occupied microclimate [80,132]. A statistically accurate personal model can still under-represent a vulnerable body or a changing activity [100,103,133]. A valid comfort vote does not identify whether the cause lies in sensing, prediction, actuation, the building, or unavailable adaptation [91,124]. Reviews of real-world personal-model control likewise show that predictive performance must be separated from deployment, realised comfort, and energy outcomes [134]. Measured compliance is therefore an intermediate result, whereas acceptable lived exposure requires concordant physical, modelled, and experiential evidence [91,132,134].

*Social thermal comfort: socially negotiated household thermal comfort.* While physiological and psychological mechanisms help explain individual responses, shared homes convert individual thermal variability into a coordination problem. Household members can differ in preferences, health, activity, schedules, room use, and access to heating controls. Comfort is consequently shaped by routines and shared expectations [11,135], as well as by vulnerability, cost, building conditions, and unequal control [136]. We use *socially negotiated household thermal comfort* descriptively, not as a new IEQ domain or standardized comfort model.

Expectations of warmth have changed with heating infrastructures, housing standards, and cultural conventions. Before central heating became widespread, households relied more on clothing, local heat, shared warm rooms, and seasonal adaptation [13,137]. Central heating subsequently helped normalise stable, spatially uniform warmth [12]. A household setpoint can therefore express care, hospitality, efficiency, frugality, or waste as well as a target temperature.

*Norms and household rules.* Household rules determine when to heat, which rooms count, who may change a thermostat, and what expenditure is acceptable. Everyday responses are shaped by climatic experience and household conventions [124,138]; high consumption may be judged wasteful, whereas restraint may be framed as responsible [139–142]. Informal arrangements can become thermostat conflicts when members have different needs or schedules [136]. ASHRAE 55 [81] and ISO 7730 [80] provide reference conditions, but they do not determine whose preference should govern a shared home. Gendered preferences [105,106], health vulnerability [143], and energy poverty [144] can therefore make apparently neutral rules distribute comfort and discomfort unequally.

*Cultural meanings.* Warmth can signify cosiness, hospitality, care, or modernity, while cooler conditions may signify freshness, efficiency, discomfort, or neglect [12]. Such meanings encourage strategies including blankets, warm clothing, fireplaces, or heating core living zones [145,146]. They can also discourage adaptation when additional heat is associated with extravagance or extra clothing with vulnerability and ageing [106,147]. Cultural meaning thus shapes both preferred conditions and which responses appear legitimate.

*Routines and socio-material practices.* Heating is embedded in waking, cooking, bathing, home working, hosting, and sleeping [148,149]. Fabric and insulation constrain system response [13,150], while emitters, thermostat location, tariffs, budgets, and user competence determine feasible adjustments [151,152]. Smart controls can coordinate schedules, feedback, and tariffs [153,154], but can also privilege the account holder, misread routines, or obscure household compromises [155–157]. They support shared comfort only when they remain responsive to household rhythms, care, vulnerability, and negotiated authority.

Thermal-comfort research may target an individual's preference [133,158]; control engineering translates it into a setpoint [159]; HCI asks whether residents can understand and correct the action [160,161]; and household research asks whose preference, account, room, schedule, and budget the system represents [136,144]. Digitisation does not remove this coordination problem: residents develop different control models [160,161], retain situated routines [162], and may remain confused by learning controls [38]. Smart heating is consequently integrated through embodied competence and domestic arrangements, not adopted as an isolated device function [163,164].

The same override can indicate prediction error, a request for agency, household compromise, care for a vulnerable resident, or an unaffordable schedule [38,128,136,144]. Evidence from the 2022 energy-price crisis shows that vulnerability and environmental concern can configure heating practices differently, although this bounded qualitative evidence is not a prevalence estimate for Europe [165]. An individually valid prediction becomes a household-level success only when the dwelling can deliver it [166,167] and the allocation of comfort, authority, and cost remains acceptable [136,168].

### 3.2. Residents' experience & practices – energy management: consumption and production

*Energy consumption.* Residents' energy consumption describes the energy used when carriers such as electricity or gas are converted into an end-use service [169,170]. Residential consumption includes space conditioning, water heating, lighting, and appliances [2–4]. In cold and temperate regions, space heating can be a major component of household demand [5,6], but a metered total alone does not establish the warmth, health protection, affordability, or household effort that the delivered energy provides.

Heating demand is not produced by equipment alone. Envelope performance, system efficiency, controls, and climate shape technical requirements, while housing conditions and economic constraints determine what households can afford to use [144,167,168]. Observed consumption also emerges through routines, budgeting, conservation practices, and competence in managing indoor conditions [13,171,172]. Occupant practices are therefore not external disturbances added to an

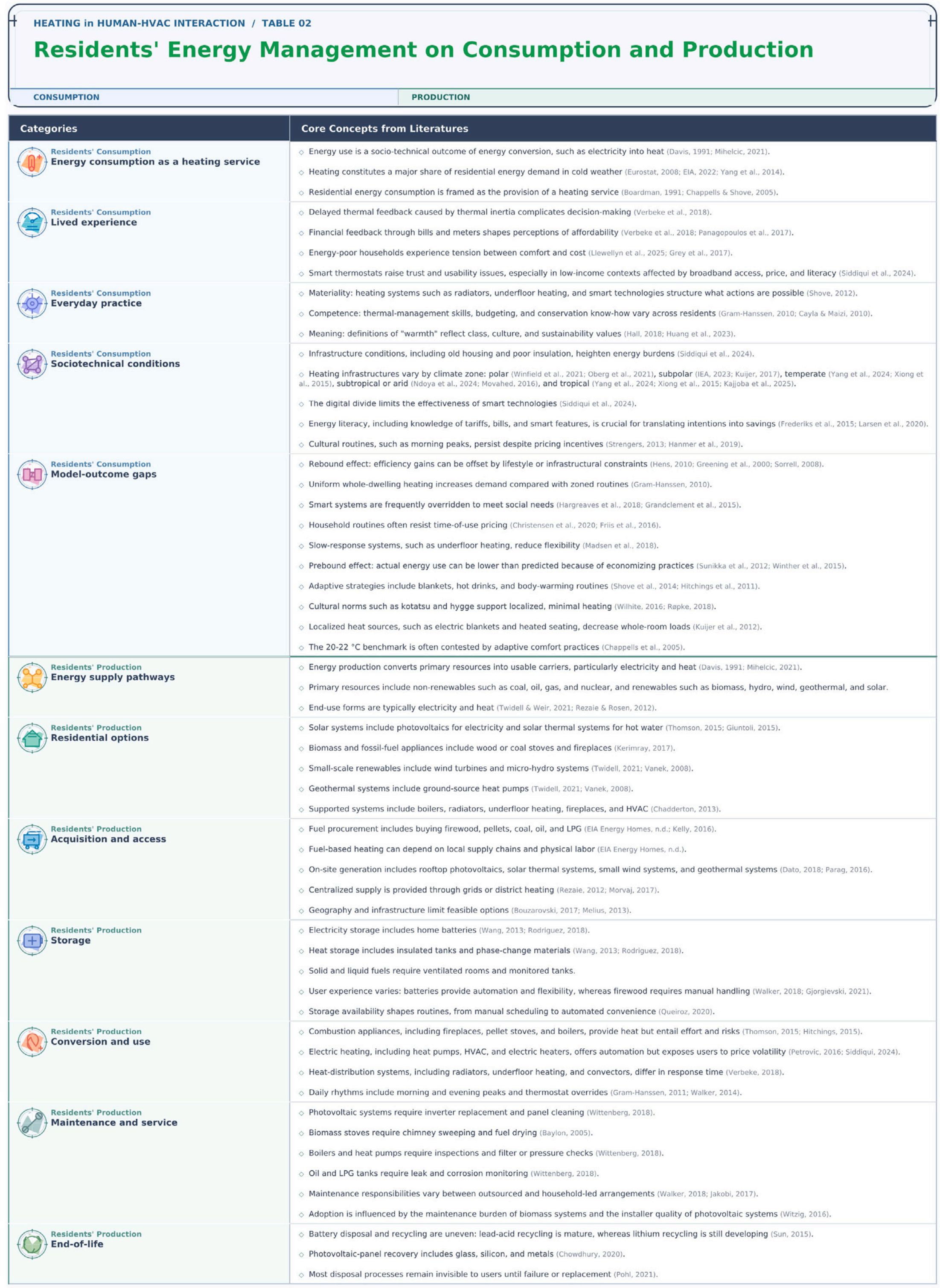

HEATING in HUMAN-HVAC INTERACTION / TABLE 02

**Residents' Energy Management on Consumption and Production**

CONSUMPTION | PRODUCTION

| Categories | Core Concepts from Literatures |
|---|---|
| Residents' Consumption<br>**Energy consumption as a heating service** | ◇ Energy use is a socio-technical outcome of energy conversion, such as electricity into heat (Davis, 1991; Mihelcic, 2021).<br>◇ Heating constitutes a major share of residential energy demand in cold weather (Eurostat, 2008; EIA, 2022; Yang et al., 2014).<br>◇ Residential energy consumption is framed as the provision of a heating service (Boardman, 1991; Chappells & Shove, 2005). |
| Residents' Consumption<br>**Lived experience** | ◇ Delayed thermal feedback caused by thermal inertia complicates decision-making (Verbeke et al., 2018).<br>◇ Financial feedback through bills and meters shapes perceptions of affordability (Verbeke et al., 2018; Panagopoulos et al., 2017).<br>◇ Energy-poor households experience tension between comfort and cost (Llewellyn et al., 2025; Grey et al., 2017).<br>◇ Smart thermostats raise trust and usability issues, especially in low-income contexts affected by broadband access, price, and literacy (Siddiqui et al., 2024). |
| Residents' Consumption<br>**Everyday practice** | ◇ Materiality: heating systems such as radiators, underfloor heating, and smart technologies structure what actions are possible (Shove, 2012).<br>◇ Competence: thermal-management skills, budgeting, and conservation know-how vary across residents (Gram-Hanssen, 2010; Cayla & Maizi, 2010).<br>◇ Meaning: definitions of "warmth" reflect class, culture, and sustainability values (Hall, 2018; Huang et al., 2023). |
| Residents' Consumption<br>**Sociotechnical conditions** | ◇ Infrastructure conditions, including old housing and poor insulation, heighten energy burdens (Siddiqui et al., 2024).<br>◇ Heating infrastructures vary by climate zone: **polar** (Winfield et al., 2021; Oberg et al., 2021), **subpolar** (IEA, 2023; Kuijer, 2017), **temperate** (Yang et al., 2024; Xiong et al., 2015), **subtropical or arid** (Ndoya et al., 2024; Movahed, 2016), **and tropical** (Yang et al., 2024; Xiong et al., 2015; Kajjoba et al., 2025).<br>◇ The digital divide limits the effectiveness of smart technologies (Siddiqui et al., 2024).<br>◇ Energy literacy, including knowledge of tariffs, bills, and smart features, is crucial for translating intentions into savings (Frederiks et al., 2015; Larsen et al., 2020).<br>◇ Cultural routines, such as morning peaks, persist despite pricing incentives (Strengers, 2013; Hanmer et al., 2019). |
| Residents' Consumption<br>**Model-outcome gaps** | ◇ Rebound effect: efficiency gains can be offset by lifestyle or infrastructural constraints (Hens, 2010; Greening et al., 2000; Sorrell, 2008).<br>◇ Uniform whole-dwelling heating increases demand compared with zoned routines (Gram-Hanssen, 2010).<br>◇ Smart systems are frequently overridden to meet social needs (Hargreaves et al., 2018; Grandclement et al., 2015).<br>◇ Household routines often resist time-of-use pricing (Christensen et al., 2020; Friis et al., 2016).<br>◇ Slow-response systems, such as underfloor heating, reduce flexibility (Madsen et al., 2018).<br>◇ Prebound effect: actual energy use can be lower than predicted because of economizing practices (Sunikka et al., 2012; Winther et al., 2015).<br>◇ Adaptive strategies include blankets, hot drinks, and body-warming routines (Shove et al., 2014; Hitchings et al., 2011).<br>◇ Cultural norms such as kotatsu and hygge support localized, minimal heating (Wilhite, 2016; Røpke, 2018).<br>◇ Localized heat sources, such as electric blankets and heated seating, decrease whole-room loads (Kuijer et al., 2012).<br>◇ The 20-22 °C benchmark is often contested by adaptive comfort practices (Chappells et al., 2005). |
| Residents' Production<br>**Energy supply pathways** | ◇ Energy production converts primary resources into usable carriers, particularly electricity and heat (Davis, 1991; Mihelcic, 2021).<br>◇ Primary resources include non-renewables such as coal, oil, gas, and nuclear, and renewables such as biomass, hydro, wind, geothermal, and solar.<br>◇ End-use forms are typically electricity and heat (Twidell & Weir, 2021; Rezaie & Rosen, 2012). |
| Residents' Production<br>**Residential options** | ◇ Solar systems include photovoltaics for electricity and solar thermal systems for hot water (Thomson, 2015; Giuntoli, 2015).<br>◇ Biomass and fossil-fuel appliances include wood or coal stoves and fireplaces (Kerimray, 2017).<br>◇ Small-scale renewables include wind turbines and micro-hydro systems (Twidell, 2021; Vanek, 2008).<br>◇ Geothermal systems include ground-source heat pumps (Twidell, 2021; Vanek, 2008).<br>◇ Supported systems include boilers, radiators, underfloor heating, fireplaces, and HVAC (Chadderton, 2013). |
| Residents' Production<br>**Acquisition and access** | ◇ Fuel procurement includes buying firewood, pellets, coal, oil, and LPG (EIA Energy Homes, n.d.; Kelly, 2016).<br>◇ Fuel-based heating can depend on local supply chains and physical labor (EIA Energy Homes, n.d.).<br>◇ On-site generation includes rooftop photovoltaics, solar thermal systems, small wind systems, and geothermal systems (Dato, 2018; Parag, 2016).<br>◇ Centralized supply is provided through grids or district heating (Rezaie, 2012; Morvaj, 2017).<br>◇ Geography and infrastructure limit feasible options (Bouzarovski, 2017; Melius, 2013). |
| Residents' Production<br>**Storage** | ◇ Electricity storage includes home batteries (Wang, 2013; Rodriguez, 2018).<br>◇ Heat storage includes insulated tanks and phase-change materials (Wang, 2013; Rodriguez, 2018).<br>◇ Solid and liquid fuels require ventilated rooms and monitored tanks.<br>◇ User experience varies: batteries provide automation and flexibility, whereas firewood requires manual handling (Walker, 2018; Gjorgievski, 2021).<br>◇ Storage availability shapes routines, from manual scheduling to automated convenience (Queiroz, 2020). |
| Residents' Production<br>**Conversion and use** | ◇ Combustion appliances, including fireplaces, pellet stoves, and boilers, provide heat but entail effort and risks (Thomson, 2015; Hitchings, 2015).<br>◇ Electric heating, including heat pumps, HVAC, and electric heaters, offers automation but exposes users to price volatility (Petrovic, 2016; Siddiqui, 2024).<br>◇ Heat-distribution systems, including radiators, underfloor heating, and convectors, differ in response time (Verbeke, 2018).<br>◇ Daily rhythms include morning and evening peaks and thermostat overrides (Gram-Hanssen, 2011; Walker, 2014). |
| Residents' Production<br>**Maintenance and service** | ◇ Photovoltaic systems require inverter replacement and panel cleaning (Wittenberg, 2018).<br>◇ Biomass stoves require chimney sweeping and fuel drying (Baylon, 2005).<br>◇ Boilers and heat pumps require inspections and filter or pressure checks (Wittenberg, 2018).<br>◇ Oil and LPG tanks require leak and corrosion monitoring (Wittenberg, 2018).<br>◇ Maintenance responsibilities vary between outsourced and household-led arrangements (Walker, 2018; Jakobi, 2017).<br>◇ Adoption is influenced by the maintenance burden of biomass systems and the installer quality of photovoltaic systems (Witzig, 2016). |
| Residents' Production<br>**End-of-life** | ◇ Battery disposal and recycling are uneven: lead-acid recycling is mature, whereas lithium recycling is still developing (Sun, 2015).<br>◇ Photovoltaic-panel recovery includes glass, silicon, and metals (Chowdhury, 2020).<br>◇ Most disposal processes remain invisible to users until failure or replacement (Pohl, 2021). |

otherwise complete engineering model; they help constitute the load observed in operation [173–175].

Thermal inertia makes the relation between action, energy input, and comfort difficult to interpret. A thermostat adjustment can change energy use immediately while room temperature responds slowly [176]. Residents consequently rely on bills, meters, displays, schedules, and remembered system behaviour to connect control actions with comfort and cost [177]. Smart thermostats attempt to anticipate this delay through learned routines and preheating [178]; however, inaccurate occupancy or routine inference can heat empty rooms, miss occupied periods, or create unwanted temperature trajectories [14,38]. An override may therefore indicate agency or correction rather than a failure to behave efficiently.

For energy-poverty households, heating decisions involve direct trade-offs among comfort, safety, and affordability [144,167,179]. Poor insulation and inefficient equipment can raise the expenditure required to maintain healthy warmth [43,180,181], while inaccessible controls or digital exclusion can limit the benefits of automation. Smart features depend on affordability, connectivity, usability, trust, and energy literacy [38,178]. Low recorded consumption must therefore be interpreted alongside indoor conditions and household resources; otherwise, harmful underheating can be misclassified as successful conservation [182].

Rebound and prebound effects expose related model–outcome gaps. After an efficiency improvement, households may heat more rooms, raise setpoints, or extend heating periods, reducing the expected savings [44,74,75,183]. Before retrofit, inefficient homes may consume less than models predict because residents already use selective heating, lower temperatures, or other forms of economising [45,47]. Neither gap is adequately explained by irrational behaviour: rebound may represent an expanded service level, while prebound consumption may conceal prior deprivation.

Demand also has temporal, spatial, and material structure. Morning and evening peaks are tied to bathing, meals, work, school, care, and preheating, and may resist price signals when activities cannot easily be rescheduled [184–186]. Slow-response systems can favour long-duration operation, whereas other households use intermittent or room-selective heating that diverges from continuous whole-dwelling assumptions [187–190]. Clothing, blankets, hot drinks, local radiant sources, and core living zones further show that residents sometimes procure warmth by heating bodies or selected spaces rather than the complete dwelling [145,146,148,191,192].

The same low-demand observation can indicate efficient service delivery, voluntary sufficiency, selective occupancy, or harmful underheating [47,182,189]. Conversely, higher post-retrofit consumption can indicate waste, rebound, recovery from inadequate warmth, or an expanded heated area [44,45,75]. Distinguishing these explanations requires the baseline heating pattern, delivered indoor conditions, household affordability, and the spatial and temporal distribution of warmth [144,188,189]. Energy saving is therefore an intermediate outcome rather than sufficient evidence of household-level success [168,182].

*Energy production and supply.* Residents' energy production and supply, concern the generation, delivery, conversion, and storage arrangements that make usable electricity or heat available [169,170]. Residential heating may depend on grid electricity, gas, oil, biomass, solar thermal systems, photovoltaic-supported heat pumps, geothermal systems, or district-heating networks [193–197]. For residents, these arrangements are experienced through tariffs, fuel acquisition, storage, emissions, maintenance, controllability, and dependence on wider service networks.

*Acquisition and supply.* Some households procure firewood, pellets, coal, heating oil, or liquefied petroleum gas directly [198–201]. These arrangements require delivery coordination, storage space, physical labour, and seasonal planning. Other homes depend on centralized or on-site infrastructures, including electricity grids, district heating, rooftop photovoltaics, solar thermal collectors, and heat pumps [202–205]. Their feasibility varies with geography, dwelling type, roof suitability, finance, fuel-delivery networks, and infrastructure availability [206–208]. Technology comparisons therefore identify context-specific possibilities rather than a universally preferable carrier or device.

*Storage, conversion, and use.* Photovoltaic electricity can be stored in batteries, while solar heat can be retained in water tanks or thermal-storage materials; wood, pellets, oil, and LPG require different forms of physical storage and monitoring [209,210]. Batteries and thermal storage can support automation and temporal flexibility [211,212], whereas direct-fuel storage can increase manual work and dependence on deliveries [213]. Combustion-based systems may provide valued experiences of warmth and tradition, but also introduce labour, safety, and indoor or outdoor air-quality concerns [214–217]. Electric heating and heat pumps can reduce on-site combustion and support automation, while increasing exposure to electricity prices, grid conditions, connectivity, and platform control [218,219].

*Maintenance and responsibility.* The work of maintaining heating differs across technologies. Photovoltaic systems can require inverter replacement and cleaning; biomass systems require fuel management and chimney maintenance; boilers and heat pumps require inspections, filters, or pressure checks; and oil or LPG tanks require leak and corrosion monitoring [43,220,221]. Some tasks are transferred to installers or service contracts, while others remain household responsibilities [213,222]. Adoption and continued use consequently depend on installer quality, repair arrangements, user learning, and whether maintenance responsibilities are visible and manageable [223].

*End-of-life.* Heating-system components, batteries, photovoltaic panels, and associated electronics create end-of-life questions about repair, reuse, material recovery, and safe disposal [48,224–226]. These processes may remain invisible to residents until replacement or failure, but they remain part of the lifecycle consequences of residential heating [227,228].

Consumption and production meet at the energy service of attainable warmth [144,148]. An automated electric system can reduce daily manual effort while increasing tariff and platform dependence [38,178,218]; a direct-fuel system can increase local control while adding labour, combustion risk, and supply-chain dependence [214,215,217]. Likewise, nominally low consumption can represent efficient service, voluntary restraint, or inadequate heating [47,182]. Evaluating Human–HVAC Interaction therefore requires more than comparing kilowatt-hours or technologies: it requires identifying which heating service was delivered, to whom, under what indoor conditions, at what household cost, and with what health, agency, maintenance, air-quality, and lifecycle consequences [167,168,170]. The central tension is not a generic comfort–energy trade-off, but whether efficiency and decarbonisation improve attainable warmth without shifting burdens or risks onto vulnerable residents [144,168,182].

### 3.3. Heating in HVAC mechanics: indoor conditions and energy performance

#### 3.3.1. Heating mechanics – thermal conditions and other IEQ conditions

Indoor environmental quality (IEQ) comprises four established domains: thermal conditions, indoor air quality, acoustic conditions, and visual conditions [92,93,229]. Heating acts directly on the thermal environment, while interacting with indoor air quality through combustion, ventilation, infiltration, moisture, and shared HVAC equipment. It can also couple with acoustic conditions through equipment and airflow noise, and with visual conditions through solar gains, glazing, shading, and blind control. These interactions can affect one another, but the four domains retain distinct variables and assessment criteria. Accordingly, ISO 7730:2025 and ASHRAE 55–2023 are used here for thermal conditions [80,81], whereas ISO 17772-1 and EN 16798-1 provide the broader IEQ framework [92,93].

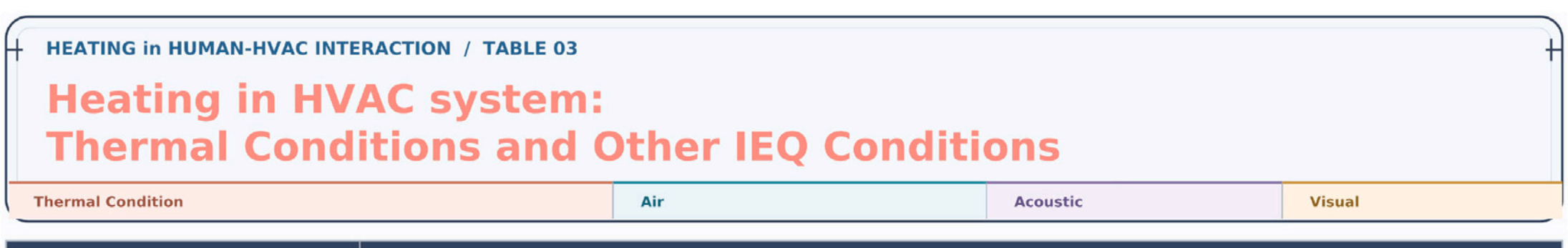

| Catagories | Concepts from Literatures |
|---|---|
| Thermal conditions | ◇ Thermal variables include air temperature, mean radiant temperature, air velocity, relative humidity, metabolic rate, and clothing insulation (ISO 2017; ASHRAE, 2024; ISO 7730, 2025).<br>◇ Heating systems are classified by energy source and heat-transfer mode (ISO 7730, 2025; ASHRAE, 2024).<br>◇ Measurement uses sensors, thermostatic radiator valves (TRVs), smart meters, and prediction models (Fanger, 1970; Yang, 2014; van Hoof, 2008).<br>◇ Measured and experienced temperature can diverge (Fanger, 1970).<br>◇ Emerging systems include predictive smart heating and heated wearables (Lee et al., 2025; Peng et al., 2020).<br>◇ Humidity affects health, material durability, and comfort (Fang, 2004; Jones, 1999), as well as heating efficiency and thermal inertia (Verbeke, 2018).<br>◇ Humidity sensing varies across heating and HVAC systems (ASHRAE, 2024).<br>◇ Residents also regulate humidity through windows, humidifiers, and plants (Wolkoff, 2018).<br>◇ Building materials and moisture dynamics shape thermal inertia (Mirrahimi et al., 2016).<br>◇ Mean radiant temperature strongly influences perceived warmth (ISO 7730, 2025).<br>◇ Radiant conditions are often unmanaged except in some smart systems (Wang et al., 2021).<br>◇ Architecture - including orientation, insulation, and passive design - modulates radiant environments (Reynolds, 2000).<br>◇ External and seasonal conditions also affect radiant environments (Tehrani et al., 2024).<br>◇ Thermal criteria remain distinct from indoor-air-quality, acoustic, and visual criteria. |
| Indoor air quality | ◇ Assessment considers pollutant sources and concentrations, including $CO_2$, VOCs, $PM_{2.5}$, and combustion products (Cincinelli et al., 2017).<br>◇ HVAC may manage ventilation and filtration, whereas stoves and radiators provide heat without active air regulation (ASHRAE, 2024).<br>◇ Window opening, purifiers, and plants shape indoor air (Liu et al., 2022).<br>◇ CO2 is a ventilation-related proxy only in defined contexts.<br>◇ Heating and ventilation can alter air exchange, moisture, and pollutant retention.<br>◇ Indoor-air-quality criteria remain distinct from thermal criteria.<br>(ISO 17772, 2017; ASHRAE, 2024; ISO 7730, 2025). |
| Acoustic conditions | ◇ Assessment considers sound levels, external noise, and reverberation.<br>◇ HVAC equipment and airflow noise belong to the acoustic domain.<br>◇ Fans, pumps, valves, ducts, and air movement can introduce noise.<br>◇ Acoustic criteria remain distinct from thermal criteria.<br>(ISO 17772, 2017). |
| Visual conditions | ◇ Assessment considers illuminance, glare, daylight, view, and circadian support (Tahkamo et al., 2019).<br>◇ Solar gains, orientation, glazing, insulation, and shading can affect radiant conditions and heating demand (Reynolds, 2000; Tehrani et al., 2024).<br>◇ Visual criteria remain distinct from thermal criteria.<br>(ISO 17772, 2017). |

*Temperature.* Temperature is the primary parameter through which heating-related HVAC performance is measured and controlled [230, 231]. Heating systems differ by energy source, control mechanisms, response time, and conductive, convective, radiative, or combined heat-transfer modes [80,231]. Thermostatic radiator valves may regulate temperature through internal mechanisms, whereas central, electric, and programmable systems use sensors, meters, actuators, and control algorithms [230,232]. A conventional room thermostat normally controls HVAC operation using measured air temperature; it does not directly measure operative temperature or thermal comfort. Determining operative temperature requires consideration of both air temperature and mean radiant temperature, while air velocity must also be considered when assessing thermal comfort [80,132].

Thermal-comfort assessment also requires mean radiant temperature, air velocity, relative humidity, metabolic rate, and clothing insulation [76,80,81]. The credibility of a temperature observation depends on calibration, sensor placement, exposure time, and whether the measurement represents vertical gradients, radiant asymmetry, and the occupied microclimate [132]. Subjective-judgement scales answer a different question by recording how an occupant evaluates the exposure [91]. Thus, a technically valid measurement and an occupant's thermal evaluation constitute complementary evidence rather than substitutes.

Smart and programmable systems extend temperature control through weather data, occupancy patterns, activity sensing, and learning-based prediction [5,40,131,233]. These functions can support comfort and efficiency, but an accurate prediction does not by itself establish that the building delivered the intended condition or that residents accepted it. Reviews of residential occupant-centric control and real-world personal comfort models therefore distinguish model performance from deployment, realised comfort, and energy outcomes [133,134,166]. Wearable and localised heating further shifts the control target from entire rooms towards bodies; its energy-saving potential must likewise be evaluated alongside exposure, acceptance, and actual operation [234,235].

*Humidity.* Relative humidity is one of the six primary variables used in thermal-comfort assessment [76,80]. Moisture conditions also affect perceived air quality, respiratory health, condensation, mould risk, building durability, and material degradation [236–240]. These roles should not be collapsed into a single measure of "humidity comfort." Sensor position and accuracy, surface temperature, ventilation, occupant moisture generation, and outdoor conditions determine whether a humidity observation indicates a cause, a consequence, or only a coincident state.

Many stoves, radiators, and electric space heaters do not measure or regulate moisture directly. Integrated HVAC systems can coordinate humidity sensing, ventilation, and air treatment [30,231,241], while residents intervene through window opening, humidifiers, dehumidifiers, and other household practices [242]. Humidity therefore links a measurable thermal variable to building moisture processes and everyday action, but it does not by itself represent the wider IEQ.

*Air quality, pollutants, and air movement.* Indoor air quality is not a single physical variable. It is assessed through pollutant sources and concentrations, particulate matter, volatile organic compounds, combustion products, and air-exchange or ventilation rates; $CO_2$ is a ventilation-related proxy only under defined source and occupancy conditions [34,92,217].

Air velocity is a distinct indoor environmental parameter. By changing convective heat exchange between the body and the surrounding air, it can alter thermal sensation and thermal comfort even when air temperature remains constant [80]. Its assessment depends on the magnitude, direction, and fluctuation of air movement at occupied locations, as well as on the measurement method and averaging period [132]. Air velocity is also related to ventilation effectiveness and airflow distribution: supply and extraction rates, diffuser arrangement, obstructions, buoyancy, and occupant actions can create draughts or stagnant zones that influence pollutant removal and perceived air quality [231,243]. Local air velocity nevertheless should not be treated as interchangeable with ventilation rate or air changes per hour; these complementary measures describe different aspects of air movement and air exchange.

Many heating devices deliver warmth without actively managing air exchange or pollutant removal. Combustion systems may introduce pollutants, while airtight or inadequately ventilated dwellings may retain $CO_2$, $PM_{2.5}$, VOCs, and moisture [34,217]. Integrated HVAC can coordinate ventilation, filtration, pollutant sensing, and heat recovery [231,243], and residents also modify air conditions through windows, doors, and air-cleaning devices [244]. Because increased ventilation can also increase heating demand, fresh air, thermal conditions, and energy use form a genuine multi-objective control problem [243,245].

*Mean radiant temperature and radiant conditions.* Mean radiant temperature characterizes radiative heat exchange between a person and the surrounding surfaces and is distinct from air temperature [132]. It is a thermal-environmental parameter and not a measure of illuminance. In typical indoor conditions, mean radiant temperature is governed predominantly by long-wave infrared exchange, with characteristic wavelengths around 10 μm, whereas illuminance concerns visible radiation at approximately 0.38–0.76 μm. The two are therefore physically distinct and require different measurement quantities and instruments. Surface temperatures, glazing, stoves, underfloor systems, and local radiant devices can change thermal sensation without an equivalent change in air temperature [246]. Conventional thermostats generally do not measure this exposure directly.

Radiant conditions connect HVAC performance with surface temperatures, insulation, glazing, and thermal mass [18]. They affect thermal comfort through radiative heat exchange and should be assessed independently of visual comfort, which is addressed below through illuminance, glare, daylight, and view.

*Acoustic and visual conditions.* Acoustic and visual conditions remain separate IEQ domains even when they interact with heating. Fans, pumps, valves, ducts, and airflow may introduce noise, while glazing and facade design influence both external-noise insulation and heat loss. Visual conditions concern illuminance, glare, daylight, and view; solar gains, glazing, blinds, and shading can simultaneously alter visual conditions and heating demand [92,93,229].

#### *3.3.2. Heating mechanics – energy efficiency performance and environmental impact*

Energy performance is not a single property of a heating technology. It includes device conversion efficiency, system performance across generation and distribution, operational performance under actual controls and use, and lifecycle consequences.

*Energy efficiency.* Energy efficiency of heating describes the useful thermal energy delivered to indoor spaces or domestic hot water relative to the energy input consumed by the system [169,170,247]. Residential outcomes depend not only on this conversion ratio but also on the energy carrier, distribution infrastructure, building envelope, control strategy, operating duration, user practices, and environmental conditions [170,231,248]. Electricity, gas, oil, biomass, and district heat may be converted by boilers, furnaces, heat pumps, radiators, underfloor heating, or ventilation-based systems; their technical ratings are therefore only one part of realised performance.

*Device efficiency* concerns how effectively an appliance converts energy into usable heat. Common metrics include heat-pump coefficient of performance (COP), annual fuel utilisation efficiency (AFUE), and seasonal ratings that account for climate and usage variability [247,249,250]. Boilers, furnaces, heat pumps, radiators, convectors, and underfloor systems differ in operating temperature, responsiveness, zoning, controllability, and comfort delivery [176,231]. Central or underfloor systems may provide stable whole-home warmth but offer limited room-level correction [251]; coal, biomass, or portable-gas devices may provide rapid local warmth while adding combustion, safety, labour, and air-quality risks [194,195].

*System efficiency* includes generation, distribution, storage, emitters, the building envelope, commissioning, and control. An efficient device can still perform poorly when heat is lost through pipes, ducts, tanks, or an inadequately insulated envelope [252]. Thermostats, zoning, scheduling, and demand-response functions influence where and when heat is delivered, while thermal inertia determines how quickly a command changes indoor conditions [176]. Energy supply also changes the interpretation of performance: heat pumps powered by lower-carbon electricity can reduce operational emissions, whereas efficient fossil-fuel combustion can carry a higher greenhouse-gas burden [253].

*Operational efficiency* is produced through the interaction of systems and residents. Thermostat settings, ventilation, schedules, window opening, occupancy, and room-selective heating can create substantial differences in demand [254,255]. Preheating, adaptive scheduling, and

HEATING in HUMAN-HVAC INTERACTION / TABLE 04

## Heating in HVAC system: Energy Efficiency Performance and Environmental Impact Lifecycle

ENERGY-EFFICIENCY PERFORMANCE | ENVIRONMENTAL IMPACT LIFECYCLE STAGES

| Categories | Core Concepts from Literatures |
|---|---|
| Energy Efficiency Performance<br>**Performance chain** | ◇ Efficiency is the ratio of useful heat delivered to total primary-energy input (Davis, 1991; Willem et al., 2017).<br>◇ Inputs include electricity, gas, oil, biomass, and district heating (Mihelcic, 2021).<br>◇ Outputs are delivered through HVAC systems.<br>◇ Efficiency strongly shapes operating costs (Allen et al., 2024). |
| Energy Efficiency Performance<br>**Device performance** | ◇ Key metrics include coefficient of performance (COP), annual fuel utilization efficiency (AFUE), and seasonal performance ratings (Tong et al., 2010; Morrison, 2004).<br>◇ Appliance types include boilers, furnaces, and heat pumps.<br>◇ Comfort modes vary across room, space, and central-heating arrangements (ASHRAE, 2024).<br>◇ Central or floor-heating systems can constrain room-level control (Mahmoud et al., 2021).<br>◇ Users often perceive high installation costs and wasted heating in unoccupied rooms (Siddiqui et al., 2024).<br>◇ Traditional heating, such as biomass, may be cheaper but raises climate and safety concerns (Giuntoli, 2015). |
| Energy Efficiency Performance<br>**System performance** | ◇ Distribution losses through pipes or ducts and storage effects reduce efficiency (ASHRAE, 2011).<br>◇ Control systems - including thermostats, zoning, and smart controllers - shape delivery quality (ASHRAE, 2011).<br>◇ Efficiency is influenced by the energy and time needed to reach and maintain temperature (Verbeke et al., 2018).<br>◇ The resource base - fossil, renewable, or hybrid - affects system performance (Michelsen et al., 2016). |
| Energy Efficiency Performance<br>**Operational and service outcomes** | ◇ User practices, including thermostat settings, ventilation, and scheduling, strongly affect outcomes (Bae et al., 2025; Pritoni et al., 2015).<br>◇ Behavioral variability can generate efficiency swings comparable to technological differences.<br>◇ Smart features enable programmable schedules and preheating or cooling (Lu et al., 2010).<br>◇ AI models increasingly infer comfort preferences and optimize heating efficiency (van Hoof, 2008; Liu et al., 2019; Almadhor et al., 2025). |
| Energy Efficiency Performance<br>**Validation and context** | ◇ Heat pumps perform better in mild climates than in extreme cold (Fairey, 2004).<br>◇ Lifecycle changes, including maintenance, degradation, and aging, reduce efficiency (Xiao et al., 2009).<br>◇ Efficiency is dynamic and shifts with time, context, and operational conditions (Lee et al., 2022). |
| Environmental Impact Lifecycle<br>**Production and embodied impacts** | ◇ Manufacturing system components, including batteries and structural materials, consumes energy and produces emissions (Tsalikis et al., 2015; Martinopoulos et al., 2014). |
| Environmental Impact Lifecycle<br>**Energy distribution** | ◇ Transmitting electricity or gas entails losses and environmental impacts.<br>◇ Efficiency upgrades help reduce these burdens (Stojkov et al., 2006). |
| Environmental Impact Lifecycle<br>**Operation** | ◇ Energy consumption is driven by heating or cooling load and by the source mix of renewable and fossil energy.<br>◇ Smart thermostats and controls enhance efficiency and comfort (Moon et al., 2011; Twidell, 2021). |
| Environmental Impact Lifecycle<br>**Maintenance** | ◇ Routine upkeep preserves system efficiency.<br>◇ Remote diagnostics can reduce emissions linked to service travel (Van et al., 2022; Brusselaers et al., 2020). |
| Environmental Impact Lifecycle<br>**End-of-life** | ◇ End-of-life processes include recycling or disposing of components and replacing obsolete units with more efficient technologies (Dodoo et al., 2009). |

occupancy sensing may reduce unnecessary operation, but their effectiveness depends on usability, trust, override, and alignment with actual routines and comfort needs [233,256,257]. Feedback and energy visualisation can support more informed operation [173], although displaying data does not guarantee that residents can interpret or act on it.

Smart Energy Management Systems (SEMS) extend operational control through sensors, meters, IoT devices, weather data, on-site production, storage, and building-automation interfaces [51,258]. Analytics and machine-learning methods can forecast demand, identify patterns, support peak shifting, coordinate battery charging, and align heating with renewable availability [259–262]. When connected to actuators, these systems can coordinate heating, ventilation, lighting, shading, and storage [263,264]. However, prediction or control accuracy is an intermediate result: credible evaluation also requires evidence that actuation occurred, the building responded, comfort was maintained, and energy or cost outcomes were realised [134,166].

Thermal Comfort Prediction Systems (TCPS) use environmental, occupancy, and sometimes physiological data to infer thermal states and guide HVAC operation [76,78,265]. Inputs may include temperature, relative humidity, air velocity, mean radiant temperature, clothing, activity, occupancy, and physiological signals [80,81]. Machine learning, transfer learning, synthetic data, and reinforcement learning are increasingly used to personalise prediction and control [266–271]. Yet a statistically accurate individual prediction may remain unsuitable for another body, a changing activity, a shared household, or a dwelling that cannot deliver the target [103,133,166].

Heating performance is also dynamic. Outdoor temperature and defrost cycles affect heat-pump efficiency [272], while maintenance, ageing, fouling, corrosion, refrigerant condition, and control drift can reduce performance over time [273,274]. Short-term demonstrations can therefore overstate stable benefits when they omit seasonal operation, commissioning, faults, household changes, and maintenance.

Engineering studies commonly report conversion efficiency, control accuracy, or simulated savings; building research adds envelope losses, emitter compatibility, commissioning, and thermal response; personal-comfort research adds individual prediction; and household research asks whether warmth was usable, affordable, and actually obtained [133,144,252]. The same technically efficient device can coexist with poor whole-system performance, high bills, inadequate warmth, or repeated overrides [167,255]. Conversely, higher post-retrofit consumption may indicate lost expected savings, rebound, a larger heated area, or recovery from previously inadequate warmth [47,74,275]. Energy performance should therefore be reported as a chain of device, system, operational, and delivered-service outcomes rather than as a single efficiency value.

*Environmental impact.* The environmental impact of HVAC, encompasses the consequences of residential heating across production, energy distribution, operation, maintenance, and end-of-life [276–278]. Relevant outcomes include greenhouse-gas emissions, air pollution, resource depletion, waste, and ecological burdens. A lifecycle perspective is necessary because operational demand is only one part of the footprint; materials, energy supply, maintenance, replacement, and disposal also contribute [263,276,279].

During *production*, impacts arise from raw-material extraction, refrigerants, electronics, batteries, compressors, heat exchangers, manufacturing, and assembly [280,281]. Heat pumps and advanced controls may reduce operational emissions in suitable contexts while introducing additional embodied impacts through complex components and supply chains.

During *energy distribution*, transmission losses, methane leakage, district-heating losses, and the electricity-generation mix affect the net footprint [282]. Smart grids, district heating, and decentralised renewable generation can lower some impacts [283], but their benefit depends on infrastructure, temporal carbon intensity, and the heating service actually displaced.

The *operation* phase depends on heating demand, system performance, control, refrigerants, and energy source. Renewable electricity, solar-assisted heat pumps, and lower-carbon district heating can reduce operational emissions, whereas coal-, oil-, and gas-based heating generally retain greater direct carbon or air-pollution burdens [196,284,285]. Smart thermostats, SEMS, and adaptive controls may reduce unnecessary heating or shift demand towards renewable availability [261,262,264,286]; these savings still require validation against actual occupancy, comfort, and system operation.

*Maintenance* preserves performance through servicing, filter replacement, refrigerant management, commissioning checks, and diagnostics. Remote diagnostics can support earlier intervention and reduce some service travel [287,288], but the environmental benefit depends on successful repair and continued efficient operation.

At *end-of-life*, refrigerants, batteries, electronics, and obsolete components require safe disposal, reuse, or material recovery [289]. Retrofit and replacement decisions must also account for baseline household practice. Deep retrofits can produce substantial savings where whole homes were previously heated to higher temperatures, but smaller savings where households already used selective or low-temperature heating [275]. Ignoring the baseline service can therefore overestimate operational reductions, misallocate retrofit resources, and extend the time required to offset embodied impacts. Environmental assessment should consequently compare lifecycle burdens with the heating service actually obtained, rather than with modelled dwelling demand alone.

### 3.4. Situated interaction dynamics – user and system-initiated control inputs and feedback outputs

This section examines how residents initiate interaction with heating-related HVAC systems and how systems respond to their actions. User-initiated interaction includes seeking information, planning operation, adjusting controls, and correcting an unwanted system response. These actions connect measured conditions and control logic with lived comfort, household routines, health needs, and energy costs.

#### 3.4.1. Situated interaction dynamics – user-initiated control inputs and feedback outputs

*User-initiated control inputs.* User-initiated control inputs are actions through which residents query or adjust heating-related HVAC systems according to their comfort needs, routines, and interpretations of indoor conditions [36,164,290–293]. Residents do not necessarily act according to the controller's formal logic. Evidence on household "folk theories" and thermostat use shows that they develop situated mental models of how heating responds [160,294]. Treating an input only as a command therefore overlooks the interpretation and household work that precede it.

***Monitoring*** concerns current, historical, or anticipated indoor conditions and energy use. Thermostats, sensors, applications, dashboards, kiosks, and wearables may expose temperature, humidity, air-quality indicators, system state, and local conditions [83,295–300]. Historical records can reveal night-time temperatures, absence periods, or recurring complaints [83,299], while forecasts and occupancy models can help residents anticipate future conditions [301,302]. The interpretation that connects a data stream to a domestic action remains part of the interaction rather than an automatic consequence of data availability [303].

Energy monitoring similarly exposes consumption, current loads, peak tariffs, or projected bills through smart meters, eco-feedback, and home energy-management systems [51,304–306]. Visibility, however, is not equivalent to intelligibility. Usefulness depends on representation, relevance, energy literacy, accessibility, and whether residents can connect a value to a likely cause and a feasible response [307–310].

***Planning*** organises heating around expected comfort, routines, absence, cost, and building response. Residents may programme schedules, night setbacks, preheating, or seasonal changes, and coordinate heating with room use, windows, doors, curtains, and ventilation [51,176,311–313]. Eco-coaching and comfort-aware thermostats can support this work through recommendations, comparisons, and what-if scenarios [314,315]. Such support assumes time, interest, and energy literacy; an immediate cold room, unexpected activity, or care need may instead require direct action [37,316].

***Adjustment and override*** are therefore analytically important forms of control. Turning the heating on or off, changing a setpoint, opening a window, or disabling a learning function can indicate discomfort, prediction error, changing occupancy, slow response, distrust, household compromise, or an affordability constraint [40,128,136,144,257,313,317]. Automation can support planning through occupancy detection, geofencing, and learned schedules [38,318–320]; it can also turn the heating off when another resident remains at home or misinterpret a temporary absence. Residents may consequently override settings,

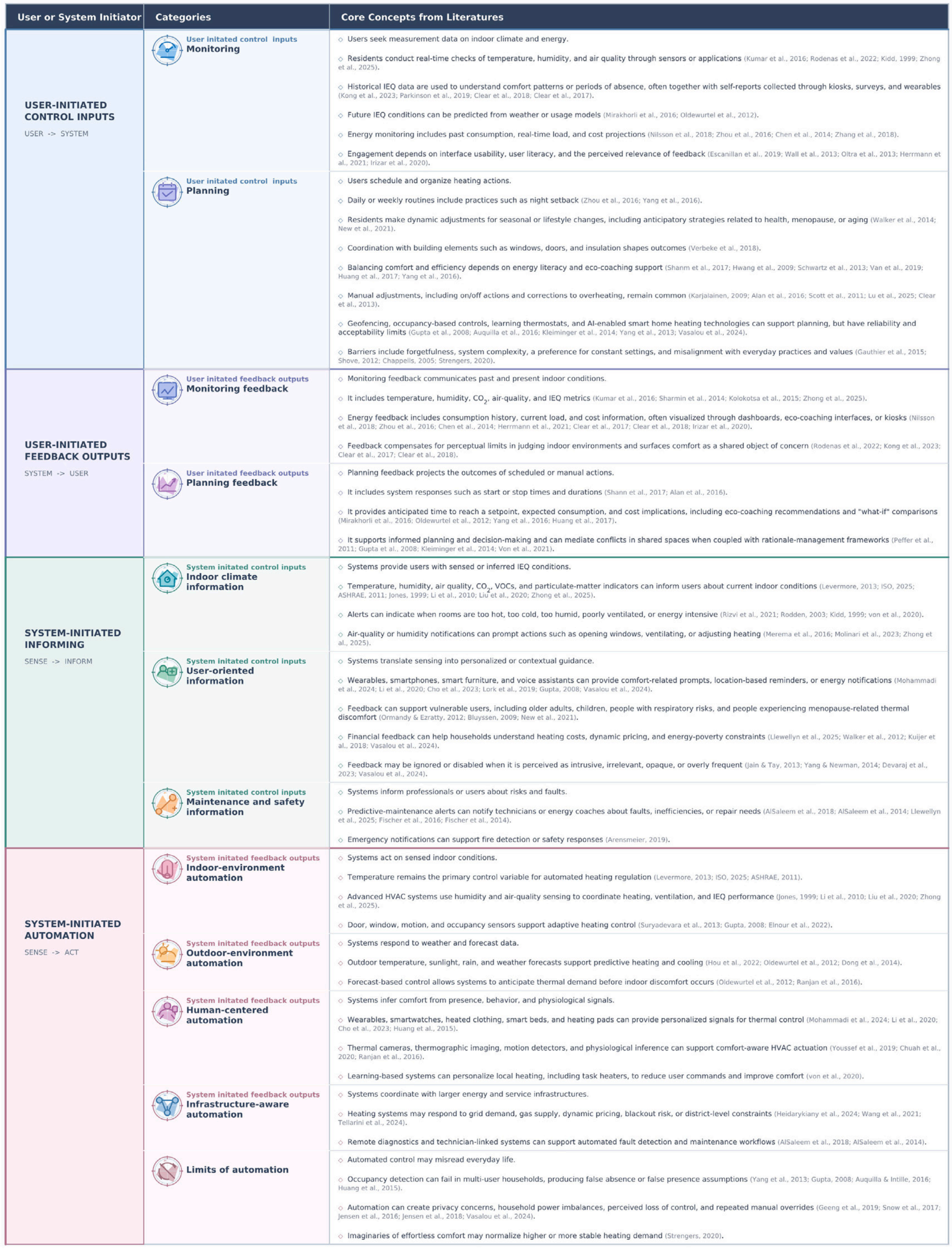

HEATING in HUMAN-HVAC INTERACTION / TABLE 05

**Situated interaction dynamics: User or System Initiated control, feedback, informing, and automation**

USER CONTROL | USER FEEDBACK | SYSTEM INFORMING | SYSTEM AUTOMATION

| User or System Initiator | Categories | Core Concepts from Literatures |
|---|---|---|
| **USER-INITIATED CONTROL INPUTS**<br>USER -> SYSTEM | User initated control inputs<br>**Monitoring** | ◇ Users seek measurement data on indoor climate and energy.<br>◇ Residents conduct real-time checks of temperature, humidity, and air quality through sensors or applications (Kumar et al., 2016; Rodenas et al., 2022; Kidd, 1999; Zhong et al., 2025).<br>◇ Historical IEQ data are used to understand comfort patterns or periods of absence, often together with self-reports collected through kiosks, surveys, and wearables (Kong et al., 2023; Parkinson et al., 2019; Clear et al., 2018; Clear et al., 2017).<br>◇ Future IEQ conditions can be predicted from weather or usage models (Mirakhorli et al., 2016; Oldewurtel et al., 2012).<br>◇ Energy monitoring includes past consumption, real-time load, and cost projections (Nilsson et al., 2018; Zhou et al., 2016; Chen et al., 2014; Zhang et al., 2018).<br>◇ Engagement depends on interface usability, user literacy, and the perceived relevance of feedback (Escanillan et al., 2019; Wall et al., 2013; Oltra et al., 2013; Herrmann et al., 2021; Irizar et al., 2020). |
| | User initated control inputs<br>**Planning** | ◇ Users schedule and organize heating actions.<br>◇ Daily or weekly routines include practices such as night setback (Zhou et al., 2016; Yang et al., 2016).<br>◇ Residents make dynamic adjustments for seasonal or lifestyle changes, including anticipatory strategies related to health, menopause, or aging (Walker et al., 2014; New et al., 2021).<br>◇ Coordination with building elements such as windows, doors, and insulation shapes outcomes (Verbeke et al., 2018).<br>◇ Balancing comfort and efficiency depends on energy literacy and eco-coaching support (Shanm et al., 2017; Hwang et al., 2009; Schwartz et al., 2013; Van et al., 2019; Huang et al., 2017; Yang et al., 2016).<br>◇ Manual adjustments, including on/off actions and corrections to overheating, remain common (Karjalainen, 2009; Alan et al., 2016; Scott et al., 2011; Lu et al., 2025; Clear et al., 2013).<br>◇ Geofencing, occupancy-based controls, learning thermostats, and AI-enabled smart home heating technologies can support planning, but have reliability and acceptability limits (Gupta et al., 2008; Auquilla et al., 2016; Kleiminger et al., 2014; Yang et al., 2013; Vasalou et al., 2024).<br>◇ Barriers include forgetfulness, system complexity, a preference for constant settings, and misalignment with everyday practices and values (Gauthier et al., 2015; Shove, 2012; Chappells, 2005; Strengers, 2020). |
| **USER-INITIATED FEEDBACK OUTPUTS**<br>SYSTEM -> USER | User initated feedback outputs<br>**Monitoring feedback** | ◇ Monitoring feedback communicates past and present indoor conditions.<br>◇ It includes temperature, humidity, $CO_2$, air-quality, and IEQ metrics (Kumar et al., 2016; Sharmin et al., 2014; Kolokotsa et al., 2015; Zhong et al., 2025).<br>◇ Energy feedback includes consumption history, current load, and cost information, often visualized through dashboards, eco-coaching interfaces, or kiosks (Nilsson et al., 2018; Zhou et al., 2016; Chen et al., 2014; Herrmann et al., 2021; Clear et al., 2017; Clear et al., 2018; Irizar et al., 2020).<br>◇ Feedback compensates for perceptual limits in judging indoor environments and surfaces comfort as a shared object of concern (Rodenas et al., 2022; Kong et al., 2023; Clear et al., 2017; Clear et al., 2018). |
| | User initated feedback outputs<br>**Planning feedback** | ◇ Planning feedback projects the outcomes of scheduled or manual actions.<br>◇ It includes system responses such as start or stop times and durations (Shann et al., 2017; Alan et al., 2016).<br>◇ It provides anticipated time to reach a setpoint, expected consumption, and cost implications, including eco-coaching recommendations and "what-if" comparisons (Mirakhorli et al., 2016; Oldewurtel et al., 2012; Yang et al., 2016; Huang et al., 2017).<br>◇ It supports informed planning and decision-making and can mediate conflicts in shared spaces when coupled with rationale-management frameworks (Peffer et al., 2011; Gupta et al., 2008; Kleiminger et al., 2014; Von et al., 2021). |
| **SYSTEM-INITIATED INFORMING**<br>SENSE -> INFORM | System initated control inputs<br>**Indoor climate information** | ◇ Systems provide users with sensed or inferred IEQ conditions.<br>◇ Temperature, humidity, air quality, $CO_2$, VOCs, and particulate-matter indicators can inform users about current indoor conditions (Levermore, 2013; ISO, 2025; ASHRAE, 2011; Jones, 1999; Li et al., 2010; Liu et al., 2020; Zhong et al., 2025).<br>◇ Alerts can indicate when rooms are too hot, too cold, too humid, poorly ventilated, or energy intensive (Rizvi et al., 2021; Rodden, 2003; Kidd, 1999; von et al., 2020).<br>◇ Air-quality or humidity notifications can prompt actions such as opening windows, ventilating, or adjusting heating (Merema et al., 2016; Molinari et al., 2023; Zhong et al., 2025). |
| | System initated control inputs<br>**User-oriented information** | ◇ Systems translate sensing into personalized or contextual guidance.<br>◇ Wearables, smartphones, smart furniture, and voice assistants can provide comfort-related prompts, location-based reminders, or energy notifications (Mohammadi et al., 2024; Li et al., 2020; Cho et al., 2023; Lork et al., 2019; Gupta, 2008; Vasalou et al., 2024).<br>◇ Feedback can support vulnerable users, including older adults, children, people with respiratory risks, and people experiencing menopause-related thermal discomfort (Ormandy & Ezratty, 2012; Bluyssen, 2009; New et al., 2021).<br>◇ Financial feedback can help households understand heating costs, dynamic pricing, and energy-poverty constraints (Llewellyn et al., 2025; Walker et al., 2012; Kuijer et al., 2018; Vasalou et al., 2024).<br>◇ Feedback may be ignored or disabled when it is perceived as intrusive, irrelevant, opaque, or overly frequent (Jain & Tay, 2013; Yang & Newman, 2014; Devaraj et al., 2023; Vasalou et al., 2024). |
| | System initated control inputs<br>**Maintenance and safety information** | ◇ Systems inform professionals or users about risks and faults.<br>◇ Predictive-maintenance alerts can notify technicians or energy coaches about faults, inefficiencies, or repair needs (AlSaleem et al., 2018; AlSaleem et al., 2014; Llewellyn et al., 2025; Fischer et al., 2016; Fischer et al., 2014).<br>◇ Emergency notifications can support fire detection or safety responses (Arensmeier, 2019). |
| **SYSTEM-INITIATED AUTOMATION**<br>SENSE -> ACT | System initated feedback outputs<br>**Indoor-environment automation** | ◇ Systems act on sensed indoor conditions.<br>◇ Temperature remains the primary control variable for automated heating regulation (Levermore, 2013; ISO, 2025; ASHRAE, 2011).<br>◇ Advanced HVAC systems use humidity and air-quality sensing to coordinate heating, ventilation, and IEQ performance (Jones, 1999; Li et al., 2010; Liu et al., 2020; Zhong et al., 2025).<br>◇ Door, window, motion, and occupancy sensors support adaptive heating control (Suryadevara et al., 2013; Gupta, 2008; Elnour et al., 2022). |
| | System initated feedback outputs<br>**Outdoor-environment automation** | ◇ Systems respond to weather and forecast data.<br>◇ Outdoor temperature, sunlight, rain, and weather forecasts support predictive heating and cooling (Hou et al., 2022; Oldewurtel et al., 2012; Dong et al., 2014).<br>◇ Forecast-based control allows systems to anticipate thermal demand before indoor discomfort occurs (Oldewurtel et al., 2012; Ranjan et al., 2016). |
| | System initated feedback outputs<br>**Human-centered automation** | ◇ Systems infer comfort from presence, behavior, and physiological signals.<br>◇ Wearables, smartwatches, heated clothing, smart beds, and heating pads can provide personalized signals for thermal control (Mohammadi et al., 2024; Li et al., 2020; Cho et al., 2023; Huang et al., 2015).<br>◇ Thermal cameras, thermographic imaging, motion detectors, and physiological inference can support comfort-aware HVAC actuation (Youssef et al., 2019; Chuah et al., 2020; Ranjan et al., 2016).<br>◇ Learning-based systems can personalize local heating, including task heaters, to reduce user commands and improve comfort (von et al., 2020). |
| | System initated feedback outputs<br>**Infrastructure-aware automation** | ◇ Systems coordinate with larger energy and service infrastructures.<br>◇ Heating systems may respond to grid demand, gas supply, dynamic pricing, blackout risk, or district-level constraints (Heidarykiany et al., 2024; Wang et al., 2021; Tellarini et al., 2024).<br>◇ Remote diagnostics and technician-linked systems can support automated fault detection and maintenance workflows (AlSaleem et al., 2018; AlSaleem et al., 2014). |
| | **Limits of automation** | ◇ Automated control may misread everyday life.<br>◇ Occupancy detection can fail in multi-user households, producing false absence or false presence assumptions (Yang et al., 2013; Gupta, 2008; Auquilla & Intille, 2016; Huang et al., 2015).<br>◇ Automation can create privacy concerns, household power imbalances, perceived loss of control, and repeated manual overrides (Geeng et al., 2019; Snow et al., 2017; Jensen et al., 2016; Jensen et al., 2018; Vasalou et al., 2024).<br>◇ Imaginaries of effortless comfort may normalize higher or more stable heating demand (Strengers, 2020). |

restrict inputs, or disable learning to correct the system and restore control [164,317,321]. Labelling these actions as inefficient without their context removes information needed to diagnose the interaction.

*User-initiated feedback outputs.* User-initiated feedback outputs are the information and system responses produced after residents monitor, schedule, adjust, or override heating controls [36,164,290,291]. Feedback should make the control loop legible by communicating the current state, the system's interpretation, the expected consequence, the time required, cost or energy implications, uncertainty, and a feasible means of correction.

***Monitoring feedback*** may report current or historical temperature, humidity, ventilation-related indicators, energy use, heating load, and system state [83,296–299]. Dashboards, kiosks, thermostats, and mobile applications can make such evidence visible. Studies in workplaces and shared buildings show that displays can also make competing comfort claims discussable [322,323]; this evidence illustrates a representation mechanism but does not by itself establish residential household outcomes.

Measurement-based feedback helps residents assess variables that cannot be estimated reliably through sensation alone. Nevertheless, a calibrated physical observation and a subjective thermal judgement answer different questions [91,132]. Numerical feedback therefore does not automatically resolve disagreement when sensor placement, bodily sensitivity, household expectations, or building response produces a different lived outcome [79,322].

***Planning feedback*** communicates the anticipated consequences of an action, including start and stop times, heating duration, time to reach a target, expected energy use, projected cost, or alternative schedules [301,302,311,313]. Eco-coaching can compare comfort and savings [314,315], while rationale displays can make the basis for a shared setting more visible [324]. When residents cannot understand why a temperature or schedule was selected, feedback may reduce trust and encourage a return to manual control [164,317,321].

Feedback should consequently be evaluated through comprehension and action rather than exposure alone. Human–building-interface research connects interface access and quality with behaviour, comfort, and energy outcomes while cautioning that these outcomes are not interchangeable [325]. A value that residents cannot interpret, a recommendation they cannot implement, or an override that does not restore acceptable operation has not closed the interaction loop. Correction success, recovery time, non-use, and the reasons for override are therefore relevant outcomes.

An override is one observable event but not one explanation. Control engineering may classify it as a deviation from an optimal policy or a prediction error [38]; thermal-comfort research may interpret it as evidence of discomfort [128]; HCI may identify a repair of opaque automation [317]; household research may reveal competing schedules or control rights [136]; and energy-poverty research may identify an attempt to contain costs [144]. These explanations can coexist. Event logs establish what changed and when, but distinguishing the explanations requires time-linked indoor conditions, building response, system rationale, household context, affordability, and residents' accounts. User control is therefore not merely a disturbance to automation; it is a diagnostic channel through which discrepancies become visible.

#### *3.4.2. Situated interaction dynamics – system-initiated informing and automation*

System-initiated interaction begins when a heating system senses or infers a condition and responds without a resident first requesting information. The response may preserve a decision point by informing residents or service actors, or it may directly change the environment through automation. Both follow an evidence chain—sensing, inference, decision, actuation, explanation, building response, and experienced outcome—but automation carries a greater burden because an erroneous inference is translated directly into environmental change.

*Informing users and service actors.* System-initiated informing includes alerts, recommendations, explanations, and maintenance messages. Systems may report cold or hot conditions, humidity, ventilation-related indicators, open windows, high energy use, dynamic prices, projected cost, comfort risk, or a detected fault [179,300,326–328]. These messages differ from user-initiated monitoring because the system decides that a condition is sufficiently relevant to interrupt or prompt action.

Wearables, phones, smart clothing, furniture, and voice interfaces can deliver location- or person-related prompts [329–332], while technician-facing systems can communicate diagnostic and maintenance needs [333–335]. Energy platforms may also notify households of high consumption, peak demand, or projected costs [14,179,208]. The value of informing depends on timing, priority, explanation, accessibility, and feasible action; frequent, intrusive, irrelevant, or unaffordable recommendations may be ignored or disabled [41,164,336,337].

*Automation.* Automation uses sensed or inferred data to change heating, ventilation, filtration, zoning, or schedules without immediate user input. Environmental sensors can regulate temperature and coordinate heating with moisture or pollutant control, while door, window, motion, and occupancy sensing can reduce output during absence or respond to open windows [237,252,327,338,339]. Weather forecasts can support preheating or delayed operation where thermal inertia makes purely reactive control too slow [302,340,341].

Personalised automation adds environmental sensing, thermography, wearables, or physiological signals to infer thermal state and operate local or central systems [329,331,342,343]. These approaches can reduce repeated commands and respond to individual variability, but wearable technologies are not yet sufficiently reliable to serve as stand-alone measures of thermal comfort. Many devices sample skin temperature at only one body location, which may not represent mean skin temperature or the whole-body thermal state, particularly in non-uniform environments or during localised heating or cooling [78,344,345]. Calibration, population validity, wearing burden, and data governance introduce further uncertainty. A physiological signal is evidence about a measured response; it is not a complete statement of preference, household acceptability, or safe control.

Automation can also coordinate heating with tariffs, demand response, grid constraints, storage, district systems, and remote service infrastructures [43,346,347]. A technically optimal grid response may nevertheless shift thermal risk, cost, or inconvenience to residents with limited flexibility. Fault detection and remote diagnostics can improve maintenance, but only when responsibility, service access, fallback, and recovery are clear [333,334].

The recurrent failures occur at the interfaces between the evidence-chain links. A point sensor may miss an occupied microclimate; occupancy inference may confuse inactivity with absence; a single account holder may be treated as the household; and a correct digital command may fail because of slow thermal response, incompatible emitters, or poor connectivity [38,318,319,342]. Geofencing, location, physiological, and behavioural data also create privacy and permission concerns, while unequal application ownership can divide households into controlling and passive users [157,348]. Residents respond by overriding, restricting data, or disabling automation when a system is opaque, intrusive, or thermally unacceptable [164,317,349].

Automation can remove repetitive adjustments while creating new work in configuration, interpretation, exception handling, privacy management, troubleshooting, and service coordination. Its success cannot therefore be inferred from prediction accuracy, command execution, or short-term energy savings alone. It requires evidence that sensing was valid, the inference represented the relevant resident or household, actuation produced the intended building response, the action remained understandable and contestable, and acceptable comfort, health, cost, and energy outcomes were sustained [166,325,350]. Non-use may indicate poor usability, but it can also be a rational response to inaccurate inference, intrusive data collection, unequal household authority, unaffordable operation, or unavailable repair. The central tension is therefore not automation versus manual control in the abstract, but how automation can reduce routine work while preserving agency, safe fallback, and the capacity to diagnose and correct failure.

Many studies validate only one link in this interaction chain, such as sensor accuracy, prediction performance, interface exposure, or simulated energy savings. Fewer combine longitudinal control logs, calibrated environmental measurements, building response, explanations shown to residents, accounts from all affected household members, and realised comfort, health, cost, and energy outcomes [166,350]. This missing cross-link evidence limits claims that a technically functional smart-heating intervention is successful at the household level.

## 4. Discussion

This Discussion answers the two research questions and clarifies the contribution of the Human–HVAC Interaction framework for building researchers, designers, engineers, housing actors, and service providers. RQ1 conceptualises residential heating through three interdependent dimensions—*Situated Interaction Dynamics*, *Residents' Experience and Practices*, and *HVAC System Mechanics*—connected through a recurring interaction cycle. RQ2 uses this account to examine three coupled objectives—healthy thermal conditions within indoor environmental quality (IEQ), affordability, and sustainability—and four tensions that can prevent a technically functional intervention from becoming a successful household outcome. For Building and Environment readers, the framework provides a diagnostic and design structure for locating where an intervention succeeds or fails across measurement, modelling, actuation, building response, bodily experience, household negotiation, correction, and maintenance. It therefore complements occupant-centric control and human–building-interface research by connecting technical performance with the social organisation and practical sustainability of residential heating [166,325,351].

### 4.1. RQ1: Conceptualizing heating in human–HVAC interaction

The answer to RQ1 is that heating in Human–HVAC Interaction is not simply temperature control or an interface between one user and one thermostat. It is the situated coordination of physical conditions, computational representations, building response, embodied experience, household organisation, and energy-service arrangements. The three dimensions distinguish how interaction occurs, how its consequences are lived, and what material and computational processes make heating possible. No dimension alone explains whether heating works for a particular resident, household, or dwelling.

#### 4.1.1. Three constitutive dimensions

*Residents' experience and practices.* Residents' Experience and Practices concern how bodies encounter heating, how residents interpret it, and how it becomes organised through routines and household negotiation. Thermal experience varies with air and mean radiant temperature, humidity, air movement, clothing, activity, physiology, health, expectation, and adaptive capacity [76,77,94,95,113,124]. Residents respond through clothing, window opening, room selection, selective heating, schedules, curtains, local warming, and overrides. These practices are shaped by culture, care, material arrangements, prices, energy literacy, and control authority [11,40,79,135,163,187,352–355].

Engineering and modelling studies may represent these actions as behavioural variation around expected operation; social-practice and household studies show that they can instead constitute how warmth is produced and shared [6,13,153,356,357]. Our position is that individual measurement is necessary but insufficient for technological support in an inherently social setting. A valid personal preference cannot establish whose preference should govern a shared room, who can access the application, who performs configuration and repair, or whether another resident's discomfort remains invisible. Support for household heating should consequently include shared or differentiated permissions, multiple forms of feedback, negotiation, care-sensitive exceptions, and ways to represent residents who are not the account holder.

*HVAC system mechanics.* HVAC System Mechanics comprises how heating-related conditions are measured, predicted, produced, distributed, and maintained: sensing; comfort, demand, and control models; plant, emitters, and actuators; ventilation; envelope performance; thermal mass; zoning; commissioning; and maintenance. Heating acts directly on thermal conditions and can couple with indoor air quality through ventilation, combustion, infiltration, filtration, and moisture; it should not be used as a synonym for the full multidimensional concept of IEQ. Standards and models such as PMV, adaptive comfort, ASHRAE 55, and ISO 7730 support thermal measurement, prediction, comparison, and control [76,80,81,94,358], while ISO 17772-1 and EN 16798-1 provide a broader IEQ framing [92,93].

Occupied dwellings test technical assumptions through heterogeneous rooms, sensor placement, slow response, incompatible components, changing activities, weather, faults, and maintenance. Mechanics also include efficiency, operating cost, maintenance burden, and life-cycle effects, which determine whether a nominally capable system remains usable and sustainable [18,166,265]. A reached setpoint or executed command is therefore an intermediate result. Technological support must connect measurements and model outputs to actuator state, building response, uncertainty, diagnostic evidence, and feasible repair. Together, HVAC System Mechanics delimit what the dwelling can deliver; Residents' Experience and Practices establish how conditions are experienced, interpreted, and socially organised; and Situated Interaction Dynamics explain how discrepancies are communicated and corrected. This connects human–building interaction, occupant-centric control, adaptive comfort, sustainable HCI, and social-practice research without reducing one tradition to another [12,50,124].

*Situated interaction dynamics.* Situated Interaction Dynamics describes how control and feedback move between residents, heating systems, buildings, and service actors. Resident-initiated interaction includes monitoring temperature, air-quality indicators, energy use, and cost; planning schedules; adjusting setpoints; overriding automation; and reporting problems [128,175,297,318]. System-initiated interaction includes sensing or inferring conditions, recommending or applying actions, and notifying residents or professionals about underheating, overheating, ventilation-related indicators, open windows, high consumption, faults, or maintenance needs [252,326,327,333,338,346]. These interactions are distributed across devices, rooms, household members, accounts, installers, and service platforms. Timing, intelligibility, permissions, and responsibility determine who can act and who bears the consequences [26,325,351]. This differs from treating interaction as a sequence of commands. An override can correct a mistaken inference, respond to care needs, negotiate a shared room, or contain cost; repeated corrections may also reveal a model, interface, emitter, or building problem. Technological support should therefore preserve the reason, context, and outcome of an action rather than recording it only as a deviation from an optimal schedule.

#### 4.1.2. From three dimensions to a recurring interaction cycle

The three dimensions become operational through a recurring cycle: *sensing → modelling → actuation → building response → bodily experience → household negotiation → correction and maintenance*.

*Sensing* captures selected environmental, equipment, occupancy, or bodily states. *Modelling* estimates demand, presence, preference, comfort, or risk, and *actuation* applies a control decision. The *building response* depends on fabric, emitters, thermal mass, ventilation, weather, commissioning, and system condition. Residents encounter the result as *bodily experience*, which enters *household negotiation* among needs, routines, costs, and control rights. *Correction and maintenance* address immediate mismatches or material faults through settings, overrides, reconfiguration, diagnosis, repair, or service coordination, feeding back into subsequent sensing and modelling.

The unit of analysis is alignment across this cycle, not an occupant, thermostat, algorithm, interface, or building in isolation. Discrepancies can arise when a point sensor misses an occupied microclimate, a model misreads inactivity, an emitter cannot realise a command, delivered warmth remains unacceptable, household needs conflict, or an override cannot repair a material fault. Physical measurement standards and subjective-judgement scales establish different forms of validity [91,132]; neither alone identifies where a discrepancy originated.

The disciplinary positions become most useful when their limits are explicit. Thermal physiology and personal comfort models identify bodily variation but do not establish whether a particular building or control action caused the outcome [103,158]. Control engineering can optimise a defined objective but may treat residents as loads, constraints, or disturbances. HCI can reveal interpretation, configuration, agency, and repair, while leaving envelope or emitter performance under-specified [325]. Household and social-practice research explains routines, care, and negotiation but does not validate instrument accuracy or actuation [12,163]. Energy-poverty research shows that low demand can indicate constrained rather than efficient service [144,167]. Whole-cycle assessment requires these forms of evidence to be related rather than substituted for one another.

Personal comfort models may improve individual prediction, but prediction remains incomplete when residents cannot interpret, contest, or correct an action, when cohabitants are not represented, or when the dwelling cannot deliver the target [158,342,343,359,360]. Conversely, an experiential account can explain why an action matters but cannot establish whether sensing, actuation, or the building response performed as intended. This asymmetry makes collaboration necessary: systems contribute measurement, prediction, coordination, and diagnostic memory, while residents contribute situated judgement, exceptions, household context, and correction.

Energy and thermal literacy can support this collaboration when feedback relates indoor conditions and demand to routines, health, and budgets [141,361,362]. Abstract values or unexplained alerts are insufficient. Interfaces should communicate what is happening, why it matters, what evidence and uncertainty inform an action, what consequence is expected, and what correction or service route remains available. In practical terms, the RQ1 contribution is a diagnostic specification for technological support: preserve evidence across the cycle, expose the rationale for control, allow residents to contest and correct it, and retain a path from failed correction to maintenance and repair.

### 4.2. RQ2: Designing for healthy, affordable, and sustainable residential heating

The answer to RQ2 is not a single optimal setpoint, interface, or control strategy. Residential heating should be designed and evaluated against three coupled but non-interchangeable objectives: healthy thermal conditions within IEQ, affordability, and sustainability [363]. The RQ1 cycle identifies where an intervention can meet one objective while undermining another and makes the resulting trade-offs available for design rather than leaving them as unintended consequences.

#### 4.2.1. Three coupled design objectives

*Healthy thermal conditions within IEQ.* Healthy thermal conditions support safety, health, and comfort across differences in physiology, activity, exposure, age, and adaptive capacity. They are one component of IEQ rather than a synonym. Heating interacts with indoor air quality through ventilation, infiltration, combustion, humidity, condensation, filtration, and occupant action, while acoustic and visual conditions retain separate criteria. Thermal-comfort compliance alone cannot establish health or safety, especially for vulnerable residents [34,102,107,143,182,192,364–366].

Personal comfort models and physiological sensing can represent some individual variation [158,300,342,343,359], but sensing should support rather than replace user judgement. Technological support should connect measurements to likely causes, health-relevant consequences, uncertainty, and feasible actions. Healthy heating should therefore be evaluated through delivered exposure, vulnerable residents' experience, and successful correction, rather than temperature compliance alone.

*Affordability.* Affordability is the capacity to secure adequate warmth without sacrificing essential needs. It includes the energy price as well as installation, maintenance, tenure, connectivity, operating complexity, and repair or configuration labour [144,179,210,367]. Smart heating can create additional work through monitoring, correcting automation, troubleshooting, and managing service problems [164]. Low-temperature heat pumps may similarly require unfamiliar operating practices [368]. Efficiency therefore does not establish affordability when residents cannot obtain adequate warmth or sustain system operation.

Low-cost spatial, material, and embodied measures can sometimes reduce costs while maintaining comfort, including draft sealing, curtain and door practices, inexpensive sensing, layered clothing, or localised heating [40,163,369]. Such measures should not substitute for adequate building fabric, affordable energy, or whole-dwelling conditions. Recommendations should communicate expected benefit, cost, affected rooms or bodies, and health constraints, while avoiding treating harmful underheating as successful conservation.

*Sustainability.* Sustainability includes operational energy and carbon together with production, transportation, installation, maintenance, replacement, and end-of-life [48,280,286,289]. Renewable supply, storage, tariffs, demand response, district heating, cooperatives, and decentralised systems can distribute flexibility beyond individual households [370], but their value depends partly on whether residents can understand and participate in their operation. Automation may reduce routine effort while obscuring energy consequences or encouraging higher service levels, and environmental responsibility is interpreted through social and cultural as well as technical frames [371,372].

Sustainability also extends to the digital infrastructure of smart heating. Cloud analytics and AI should be justified against energy, water, privacy, compatibility, and maintenance costs. Where adequate, local processing, low-power interfaces, and minimal data architectures may be preferable. Durability, modularity, repairability, recyclability, and service accountability should therefore form part of system performance, while responsibility for decarbonisation should not be shifted entirely onto household behaviour.

#### 4.2.2. Four interdisciplinary tensions

*Tension A: Sensed and modelled conditions versus lived comfort.* Smart heating depends on partial representations. Point sensors may miss occupied microclimates; occupancy inference may confuse inactivity with absence; wearables require calibration and can burden residents; and comfort models embed assumptions about clothing, activity, exposure, and population [38,342,360]. ISO 7726 addresses physical measurement of thermal environments, whereas ISO 10,551 addresses subjective judgement [91,132]. Neither alone provides a complete account of Human–HVAC outcomes.

Sensor or model validation therefore does not validate the whole interaction cycle. Longitudinal field studies are needed to capture seasonal adaptation, household routines, maintenance, connectivity failures, and multi-occupant conflict [134,350]. Interfaces should communicate measurements together with uncertainty, likely causes, consequences, and feasible responses, allowing sensing to support situated judgement [37].

*Tension B: Individual personalisation versus household negotiation.* Personal comfort models and localised conditioning can address individual variability [158,343]. Homes, however, involve shared rooms and equipment, conflicting schedules, care responsibilities, unequal permissions, limited zoning, and common budgets [136,157,348]. An accurate individual prediction may therefore remain impossible or inequitable to implement.

Evaluation should examine the empathy among multiple users: who can change settings, whose feedback enters the model, whose discomfort remains invisible, and how compromises are reached. Occupant-centric control should represent cohabitation, authority, care, and the distribution of benefits, costs, and work. Shared permissions, multi-person feedback, room- or person-specific alternatives, and explicit conflict-resolution mechanisms are therefore important design requirements.

*Tension C: Efficiency and decarbonisation versus health and adequate warmth.* Lower setpoints and shorter heating periods can reduce demand while increasing cold or moisture risks in inefficient housing and for vulnerable residents [34,144,167,182]. Recommending more heating may likewise be unaffordable. Rebound and prebound effects show that predicted savings depend on prior service levels, comfort expectations, affordability, and post-retrofit practices [44,45,47].

Evidence from the European energy crisis since 2022 also shows varied household responses. A qualitative study of 30 Danish households found heating practices associated with vulnerability and environmental concern [165]. Rebound may occur when residents heat more rooms, for longer, or to higher temperatures after an upgrade, whereas prebound may indicate that adequate warmth was previously unaffordable [74,176,178]. Energy reduction is therefore an intermediate result. Evaluations should also report baseline heating service, delivered conditions, expenditure, vulnerability, and the distribution of consequences.

*Tension D: Automation and convenience versus agency, intelligibility, and maintenance.* Automation can reduce routine effort, anticipate thermal inertia, and coordinate heating with occupancy, weather, prices, and grid conditions [51,302,321]. It can also impose unsuitable schedules, overlook situated activities and health needs, obscure reasoning, and create work through setup, correction, troubleshooting, and service coordination [38,130,164,313,317]. Automation therefore redistributes work and authority across residents, account holders, landlords, installers, service providers, and platforms.

Collaborative systems can combine sensing and forecasting with residents' situated judgement by communicating rationale, uncertainty, expected consequences, and available corrections [166,322–324]. Eco-coaching and data-mediated negotiation can further support comfort–energy trade-offs [314,315,322,323]. Yet interfaces cannot resolve limited zoning, incompatible equipment, poor connectivity, rental restrictions, maintenance delays, or slow thermal response. Effective automation therefore requires explanation, shared permissions, correction, override, safe fallback, diagnosis, recovery, and clear service responsibility.

### 4.3. Limitations and future work

This review focuses on smart residential heating and does not comprehensively cover cooling, non-residential HVAC, clinical or industrial ventilation, or all IEQ domains. Indoor air quality is included where it directly interacts with heating through ventilation, combustion, moisture, filtration, or occupant action, while acoustic and visual conditions are not synthesised to a comparable depth. Cooling should not be treated as a symmetrical extension of heating because it involves distinct physiological and operational risks [254,373].

The RQ1 dimensions and interaction cycle may inform future studies of cooling and other built environments, but they require domain-specific validation. Cooling introduces issues such as heat stress, hydration, physiological vulnerability, different adaptive practices, and cooling-system operation [102,373–376]. Non-residential and safety-critical settings also redistribute authority, maintenance, and acceptable failure and therefore require their own standards and empirical evidence.

Future work may also examine conversational agents, augmented-reality interfaces, wearables, ambient displays, adaptive furniture, and agentic AI as ways to make HVAC conditions and controls more intelligible or to provide localised support [53,159,377–380].

## 5. Conclusions

This review synthesises 541 studies across architecture, engineering, informatics, physiology, psychology, sociology, and design to present a multidisciplinary overview of heating within Human–HVAC Interaction. It addresses fragmented residential heating research by relating residents' embodied and social experiences to the physical and energy performance of heating systems and to interactions through which systems are monitored, automated, corrected, and maintained.

In response to RQ1, the review conceptualises Heating in Human–HVAC Interaction through three interdependent dimensions: *Residents' Experience and Practices*, *Heating in HVAC System Mechanics*, and *Situated Interaction Dynamics*. Residents' Experience and Practices encompass thermal comfort and energy management, while Heating in HVAC System Mechanics encompasses thermal conditions and energy performance. Situated Interaction Dynamics connect them through resident-initiated monitoring, planning, adjustment, and correction and system-initiated sensing, informing, and automation. Together, they form a recurring cycle of *sensing, modelling, actuation, building response, bodily experience, household negotiation, correction, and maintenance*. This framing prevents the resident, interface, control model, equipment, or dwelling from being treated in isolation.

In response to RQ2, the review identifies three coupled but non-interchangeable objectives: healthy thermal conditions within IEQ, affordability, and sustainability. Healthy thermal conditions require attention to physical exposure, physiological diversity, lived experience, and heating-related interactions with indoor air quality. Affordability concerns securing adequate warmth without sacrificing essential needs and includes installation, operation, connectivity, maintenance, and household labour. Sustainability extends beyond operational efficiency to lifecycle impacts, digital infrastructure, durability, repairability, and demand-producing practices. Evidence remains non-substitutable: comfort does not establish health, low demand does not establish affordable warmth, and low operational energy does not establish low lifecycle impact. The synthesis identifies four interdisciplinary tensions: sensed and modelled conditions versus lived comfort; individual personalisation versus household negotiation; efficiency and decarbonisation versus health and adequate warmth; and automation and convenience versus agency, intelligibility, and maintenance. These explain why technically accurate measurement, prediction, or control may still produce an unacceptable household outcome. Evaluation should consider not only setpoints and energy reduction but also whether residents can understand and correct actions, whether the dwelling delivers adequate warmth, how benefits and burdens are distributed, and whether maintenance, fallback, and lifecycle responsibilities are defined.

## CRediT authorship contribution statement

**Delong Korus-Du:** Writing – review & editing, Writing – original draft, Visualization, Validation, Software, Resources, Methodology, Investigation, Formal analysis, Data curation, Conceptualization. **Gunnar Stevens:** Writing – review & editing, Writing – original draft, Validation, Supervision, Resources, Project administration, Methodology, Investigation, Funding acquisition, Formal analysis, Data curation, Conceptualization. **Alexander Boden:** Writing – review & editing, Supervision, Project administration, Methodology, Funding acquisition, Conceptualization. **Lenneke Kuijer:** Writing – review & editing, Validation, Supervision. **Apostolos K. Vavouris:** Writing – review & editing, Validation, Investigation, Formal analysis, Data curation. **Md Shajalal:** Writing – review & editing, Investigation, Formal analysis, Data curation. **Philip Engelbutzeder:** Writing – review & editing, Validation, Supervision, Formal analysis. **Omid Veisi:** Validation, Investigation, Formal analysis, Data curation. **Peter Tolmie:** Supervision, Conceptualization.

## Declaration of competing interest

The authors declare that they have no known competing financial interests or personal relationships that could have appeared to influence the work reported in this paper.

## Acknowledgements

This project has received funding from the European Union's Horizon 2020 research and innovation programme under the Marie

Skłodowska-Curie grant agreement No. 955422, and was also supported by the FUSION project, funded by the German Federal Ministry of Education and Research (BMBF) programme *Innovative University*.

## Appendix A. Supplementary data

Supplementary data for this article can be found online at doi:10.1016/j.buildenv.2026.115220.

## Data availability

I have shared the data link in the manuscript within the Methods section.